\documentclass[prc,aps,twocolumn,preprintnumbers,superscriptaddress]{revtex4}

\usepackage{mathrsfs}
\usepackage{amssymb}
\usepackage{graphicx}
\usepackage{dcolumn}
\usepackage{bm}
\usepackage{graphicx}
\usepackage{float}
\usepackage{longtable}
\usepackage{amsmath}
\usepackage{color}
\usepackage{diagbox}
\usepackage{multirow}
\usepackage{braket}
\usepackage{booktabs}
\usepackage[colorlinks=true,linkcolor=blue,filecolor=blue, urlcolor=blue,citecolor=blue]{hyperref}
\usepackage{threeparttable}
\usepackage{amsmath}

\newfont{\largemi}{cmmi10}
\newfont{\smallmi}{cmmi6}

\draft

\def\eqref#1{Eq.~(\ref{eq:#1})}

\fontsize{20}{20}

\begin{document}
	
\title{Nucleon-pair truncation of the shell model for medium-heavy nuclei}
	
\author{Y. X. Yu}
\affiliation{School of Physics Science and Engineering, Tongji University, Shanghai 200092, China}
	
\author{Y. Lu}
\affiliation{College of Physics and Engineering, Qufu Normal University} 
	
\author{G. J. Fu}
\email{gjfu@tongji.edu.cn}
\affiliation{School of Physics Science and Engineering, Tongji University, Shanghai 200092, China}
	
\author{Calvin W. Johnson}
\affiliation{Department of Physics, San Diego State University, 5500 Campanile Drive, San Diego, CA 92182-1233}

\author{Z. Z. Ren}
\affiliation{School of Physics Science and Engineering, Tongji University, Shanghai 200092, China}
	
\date{\today}

\begin{abstract}


{\bf Background:}  Computationally tractable models of atomic nuclei is a long-time goal of nuclear structure physics.  
A flexible framework which easily includes excited states and many-body correlations is the configuration-interaction shell model (SM), 
but the exponential growth of the basis means one needs an efficient truncation scheme, ideally one that includes both 
deformation and pairing correlations.


{\bf Purpose:} We propose an efficient truncation scheme of the SM: starting from a pair condensate variationally defined by Hartree-Fock single-particle states and the particle-number 
conserved Bardeen-Cooper-Schrieffer (NBCS) approximation, we carry out projection of states with good angular momentum.


{\bf Methods:} After generating Hartree-Fock single-particle states with Kramers degeneracy in a SM space, we optimize the pair amplitudes in the NBCS by minimizing the energy, and then  use linear algebra projection (LAP) of  states with good angular momentum. Both NBCS and LAP are computationally fast.

{\bf Results:} Our calculations yield good agreement with full configuration-interaction SM calculations for low-lying states of transitional and rotational nuclei with axially symmetric and triaxial deformation in medium- and heavy-mass regions: $^{44,46,48}$Ti, $^{48,50}$Cr, $^{52}$Fe, $^{60,62,64}$Zn, $^{66,68}$Ge, $^{68}$Se, and $^{108,110}$Xe.
We predict low-lying states of $^{112-114}\textrm{Ba}$ and $^{116-120}\textrm{Ce}$, nuclei difficult to reach by  large-scale SM calculations.

{\bf Conclusions:} Both pair correlation and the configuration mixing between  different intrinsic states play a key role in reproducing  collectivity and  shape coexistence, demonstrating the utility of this 
 truncation scheme of the SM to study transitional and deformed nuclei.

\end{abstract}
	
\vspace{0.4in}
	
\maketitle

\section{INTRODUCTION}

The spherical nuclear shell model (SM) is a flexible and useful framework for configuration-interaction calculations nuclear structure theory~\cite{talmibook,sm-brown,sm-talmi,sm-caurier}.
In a given single-particle basis, the so-called full configuration-interaction (FCI) considers all possible configurations and efficiently generates low-lying states, including complex 
states with multi-particle correlations. 
Yet for many heavy rotational nuclei, FCI dimensions go far beyond the current computational limit of $\sim 2-3 \times 10^{10}$.
Hence
the hunt for good truncation schemes that nonetheless incorporate important correlations--in particular, deformation 
and pairing--is of great importance.

Thus the motivation for the pair truncation of the SM.
For example, the (generalized) seniority scheme \cite{racah1942,racah1943,talmi1,talmi2,talmibook} and broken pair model \cite{brokenpair1,brokenpair2,brokenpair3}, which work very well for nearly spherical nuclei. 
The dominant building blocks are nucleon pairs coupled to angular momentum zero (denoted by $S$ pairs), due to the strong monopole pairing interaction.
The ground state of a semimagic even-even nucleus can be described by an $S$ pair condensate, and the low-lying excited states are interpreted as breaking of the $S$ pairs.
In the interacting boson model \cite{ibm1,ibm2,ibm3,ibm4}, the spin-zero $S$ pair and spin-two ($D$) pair are mapped to $s$ and $d$ bosons,  key ingredients in collective states of the vibrational, rotational, and $\gamma$-soft nuclei.
A fully fermionic treatment is found
in the nucleon-pair approximation (NPA) of the SM \cite{npa3,npaodd}, built by nucleon pairs with spins zero, two, four (denoted by $G$), six (denoted by $I$), etc. 
NPA calculations successfully reproduce low-lying states in nuclei from spherical, via transitional, and finally to deformed regions \cite{npareports,npadeformed3}, and the important role played by the $G$ (and sometimes $I$) pair in well-deformed nuclei has been demonstrated \cite{npadeformed1,npadeformed2}. 
An important ingredient for deformed and transitional nuclides was the extraction of good pairs from Hartree-Fock states \cite{npadeformed3,npadeformed1}.
Although NPA configuration spaces are much smaller than full SM ones, and despite recent significant speed-ups in 
codes driven largely by going from a $J$-coupled scheme to an $M$ or $J_z$-scheme \cite{npam1,npam2}, 
the computation with high-spin $G$ and $I$ pairs is still too burdensome for most of deformed nuclei across the nuclear chart.

On the other hand, one frequently finds in nuclear physics powerful deformed-mean-field calculations incorporating pairing, such as 
 Bardeen-Cooper-Schrieffer (BCS)~\cite{bcs1,bcs2,bcs-bohr,bcs-migdal} and  Hartree-Fock-Bogoliubov (HFB)~\cite{bcs-belyaev,peterring} calculations, which can be understood through  Nilsson single-particle orbits~\cite{peterring,bohr1998nuclear2}. 
Calculations using deformed BCS and HFB have had great success in nuclear physics, but they  break both rotational symmetry and particle number.
The BCS or HFB vacua are actually superpositions of  states with different angular momentum quantum numbers in neighboring 
(different particle number) nuclei.
One can restore the symmetries, for example recovering exact particle number by  numerical integration over the gauge angle~\cite{peterring,Dietrich1964}, or good angular momentum by  integration over Euler angles, where, for example, 
the configuration space is constructed by angular momentum projection on quasiparticle states after solving the BCS/HFB equation in the Nilsson orbits~\cite{psm,PhysRevC.77.061305,PhysRevC.88.024328}.

Recently a very fast algorithm has been proposed for the generalized seniority scheme in deformed orbits,  a powerful tool to calculate identical particle systems in a valence space of 15 major shells \cite{PhysRevC.96.034313,PhysRevC.99.014302}.
The generalized seniority scheme has good particle numbers naturally.
As this formalism constructs pairs between time-reversal orbits as in the BCS theory, we call it the particle number conserved BCS (NBCS) throughout this paper.

In this work, we extend the NBCS to open-shell nuclei, add angular momentum projection (denoted by PNBCS), and perform calculations for nuclei with axially symmetric deformation, triaxial deformation, and shape coexistence, in SM spaces with effective interactions.
In other words, the PNBCS can be regarded as a pair truncation scheme of the SM.
We apply the PNBCS to the study of nuclei with axially symmetric deformation, triaxial deformation, and shape coexistence in the medium- and heavy-mass regions.

There exist alternative approaches which restore both rotational symmetry and particle number conservation, such as the projected HFB  \cite{SCHMID2004565,PhysRevC.92.064310} and the projected general pair condensate \cite{PhysRevC.105.034317}.
Yet such computations are time consuming.
For example, Ref. \cite{PhysRevC.92.064310} showed that the variation after projection for the HFB can be only carried out in the $sd$ shell, because of the very heavy multi-dimension integration.
In Ref. \cite{lu2022nuclear}, from a random start it took hours to optimize a general pair condensate before angular momentum projection for open-shell nuclei in the $A \sim 130$ region.
On the other hand, the variation of the NBCS before angular momentum projection used in this work is very fast, even  compared with the already fast HF calculation. 
It therefore provides an alternative approach for variation after projection study, which we leave to future work.


This paper is organized as follows.
In Sec. II we introduce the framework of the PNBCS, including the HF with Kramers degeneracy, the NBCS, and the linear algebraic approach of angular momentum projection \cite{PHF1}.
In Secs. III and IV we compare the results of PNBCS with full SM and the projected Hartree-Fock, and show that the PNBCS provides us with good descriptions for low-lying states of transitional and rotational nuclei in the $pf$, $pf_{5/2}g_{9/2}$, and $sdg_{7/2}h_{11/2}$ shells.
In particular, since in the absence of pairing PNBCS would be the same as projected HF, our results show the improvement through the inclusion of pairing correlations in our wave functions.

\section{Framework}

In this paper we use Latin letters $a$, $b$, $c \ldots$ to denote SM single-particle states labeled by good quantum numbers $n$ (radial nodes), $l$ (orbital angular momentum), $j$ (total angular momentum), and  $m$ ($z$-component of angular momentum). 
We write the creation operator of a nucleon as
\begin{equation}
  \hat{c}_{a}^{\dagger}  \equiv \hat{c}_{n_{a}l_{a}j_{a}m_{a}}^{\dagger},
\end{equation}
and its time-reversed partner as
\begin{equation}
  \hat{c}_{\tilde{a}}^{\dagger} \equiv (-)^{j_{a} -m_{a} }  \hat{c}_{n_{a}l_{a}j_{a} -m_{a}}^{\dagger}.
\end{equation} 
We use  Greek letters $\alpha$, $\beta$, $\gamma \ldots$ to denote HF single-particle states, and we write the creation operator as $ \hat{a}_{\alpha}^{\dagger}$.

\subsection{HF and NBCS}

We start with the HF calculation in a shell model (SM) basis with Kramers degeneracy \cite{kramers}, that is, our HF calculation always produces degenerate time reversal single particle partners without enforcing additional constrains such as shape, orientation, parity.
%
A HF single particle state from our calculations can be written as a transformation of the original SM single particle states:
\begin{eqnarray}
\hat{a}_{\alpha}^{\dagger} = \sum_{a} U_{ \alpha a } \hat{c}_{a}^{\dagger},
\end{eqnarray}
and its time-reversed partner can be written by
\begin{eqnarray}
\hat{a}_{\tilde{\alpha}}^{\dagger} = \sum_{a}  U_{ \alpha a} \hat{c}_{\tilde{a}}^{\dagger}.
\end{eqnarray}

While details of particle number conserved BCS (NBCS) can be found in \cite{PhysRevC.99.014302,PhysRevC.96.034313}, here we brief the formulae we use in this work. 
The building blocks are collective pairs in the HF basis, i.e.,
\begin{equation}\label{3}
\hat{P}_{\tau}^\dagger\equiv \sum_{\alpha \in O_\tau} v_{\tau,\alpha} \hat{P}^\dagger_{\tau,\alpha} = \frac{1}{2} \sum_{\alpha} v_{\tau,\alpha}  \hat{a}^\dagger_{\tau,\alpha} \hat{a}^\dagger_{\tau, \tilde{\alpha}} ,
\end{equation}
where 
\begin{equation}
\hat{P}_{\tau,\alpha}^\dagger \equiv \hat{a}^\dagger_{\tau,\alpha} \hat{a}^\dagger_{\tau, \tilde{\alpha}} = \hat{P}_{\tau,\tilde{\alpha}}^\dagger,
\end{equation}
Here $\tau=\pi$ for valence protons and $\tau=\nu$ for valence neutrons;
$O_\tau$ denotes the indices that picks only one of each degenerate time-reversed pair $\alpha $ and $ \tilde{\alpha}$;
$v_{\tau,\alpha}$ is the pair structure coefficient.
We emphasize that prior work showed that using pairs extracted from a deformed HF state provides an important 
step towards the good description of rotational bands ~\cite{npadeformed3,npadeformed1,npadeformed2}.

In the NBCS, the ground state of $2N$ valence protons (or neutrons) is assumed to be an $N$-pair condensate:
\begin{equation}\label{2.a.1}
\ket{\phi_{\tau,N}}=\frac{1}{\sqrt{\chi_{\tau,N}}} (\hat{P}_{\tau}^\dagger)^N \ket{0},
\end{equation}
where $\chi_{\tau,N}$ is the normalization factor.
In order to facilitate a fast method for computing the NBCS energy, Refs. \cite{PhysRevC.99.014302,PhysRevC.96.034313,jia2015practical} introduced the $\alpha$-orbit blocked $N$-pair condensate:
\begin{equation}
\begin{aligned}
\ket{\phi^{[\alpha]}_{\tau,N}} &\equiv \frac{1}{\sqrt{\chi^{[\alpha]}_{\tau,N}}} ( \sum_{\substack{\beta \in O_\tau \\ \beta \neq \alpha}  } v_{\tau,\beta} \hat{P}^\dagger_{\tau,\beta} )^N \ket{0}  \\
& = \frac{1}{\sqrt{\chi^{[\alpha]}_{\tau,N}}} (\hat{P}_{\tau}^\dagger - v_{\tau,\alpha} \hat{P}_{\tau,\alpha}^\dagger )^N \ket{0},
\end{aligned}
\end{equation}
the effect of which is to remove 
 the pair of single-particle states $\alpha$ and $\tilde{\alpha}$.
Similarly, the $\alpha\beta$- and $\alpha\beta\gamma$-orbit blocked $N$-pair condensates are defined by 
\begin{eqnarray}
 &&\ket{\phi^{[\alpha\beta]}_{\tau,N}} \equiv \frac{1}{\sqrt{\chi^{[\alpha\beta]}_{\tau,N}}} (\hat{P}_{\tau}^\dagger - v_{\tau,\alpha} \hat{P}_{\tau,\alpha}^\dagger  - v_{\tau,\beta} \hat{P}_{\tau,\beta}^\dagger )^N \ket{0}, \\
 &&\ket{\phi^{[\alpha\beta\gamma]}_{\tau,N}}  \equiv \frac{1}{\sqrt{\chi^{[\alpha\beta\gamma]}_{\tau,N}}} \times \nonumber\\
 && \qquad (\hat{P}_{\tau}^\dagger - v_{\tau,\alpha} \hat{P}_{\tau,\alpha}^\dagger  - v_{\tau,\beta} \hat{P}_{\tau,\beta}^\dagger - v_{\tau,\gamma} \hat{P}_{\tau,\gamma}^\dagger )^N \ket{0}, 
\end{eqnarray}
respectively, where $\alpha$, $ \beta$, and $\gamma$ are not equal to each other.
The normalization factors $\chi_{\tau,N}$, $\chi^{[\alpha]}_{\tau,N}$, $\chi^{[\alpha\beta]}_{\tau,N}$, and $\chi^{[\alpha\beta\gamma]}_{\tau,N}$ can be calculated using recursive formulae; see Eqs. (\ref{appendix_1})-(\ref{appendix_6}) in the Appendix.

\subsection{NBCS energy for identical particles}

The Hamiltonian for identical particles is written as
\begin{eqnarray}
\hat{H} &=& \hat{H}_{0 \tau}+\hat{H}_{\tau\tau} \nonumber\\
&=& \sum_{\alpha \beta  } \epsilon_{\alpha \beta} \hat{a}^\dagger_{\tau,\alpha } \hat{a}_{\tau,\beta} + \frac{1}{4}\sum_{\alpha \beta \gamma \delta }  V_{\alpha \beta  \gamma \delta} \hat{a}^\dagger_{\tau,\alpha }  \hat{a}^\dagger_{\tau,\beta } \hat{a}_{\tau,\delta }  \hat{a}_{\tau,\gamma },   \nonumber\\
\end{eqnarray}
where $ V_{\alpha \beta  \gamma \delta}$ are antisymmetrized two-body matrix elements in the HF single-particle basis.
Using Eqs. (\ref{appendix_7})-(\ref{appendix_10}) in Appendix, the energy of the $N$-pair condensate is given by
\begin{equation}\label{E1}
\begin{aligned}
E_\tau \equiv & \bra{\phi_{\tau,N}} \hat{H} \ket{\phi_{\tau,N}}\\
= & \sum_{\alpha \in O_\tau}(2 \epsilon_{\alpha \alpha}+G_{\alpha \alpha})  
     \left( 1-\frac{\chi^{[\alpha]}_{\tau,N}}{\chi_{\tau,N}} \right)\\
&  + \sum_{\alpha \beta \in O_\tau }^{\alpha \neq \beta }  
   \left[ G_{\alpha \beta }  N^2  v_{\tau,\alpha}    v_{\tau,\beta}  \frac{\chi^{[\alpha \beta]}_{\tau,N-1}}{\chi_{\tau,N}} \right. \\
&  \left. \qquad\quad  + A_{\alpha \beta}  \left( 1 - \frac {\chi^{[\alpha]}_{\tau,N} + \chi^{[\beta]}_{\tau,N} - \chi^{[\alpha \beta]}_{\tau,N}}  {\chi_{\tau,N}}  \right) \right],
\end{aligned}
\end{equation}
where 
\begin{equation}
\begin{aligned}
G_{\alpha \beta} \equiv &V_{\alpha \tilde{\alpha}   \beta \tilde{\beta} } , \\
A_{\alpha \beta} =&V_{\alpha \beta   \alpha \beta}+V_{\alpha \tilde{\beta}  \alpha \tilde{\beta}  }  .
\end{aligned}
\end{equation}

The energy of the $N$-pair condensate, $E_\tau$, varies with the pair structure coefficient $v_{\tau,\alpha}$. 
In general $v_{\tau,\alpha}$ is determined by minimizing $E_\tau$, so that
\begin{equation}
\frac{\partial E_\tau }{\partial v_{\tau,\alpha}}=0,
\end{equation}
leading to the  iterative formula
\begin{equation}\label{2.a.2}
 v_{\tau,\alpha}=  \frac{ \displaystyle -\sum_{\beta \in O_\tau}^{\beta \neq \alpha} G_{\alpha \beta } v_{\tau , \beta} \chi^{[\alpha \beta]}_{\tau,N-1} }{    (d_{\tau,\alpha}+ \bra{\phi^{[\alpha]}_{\tau,N-1}} \hat{H} \ket{\phi^{[\alpha]}_{\tau,N-1}}-E_\tau)    \chi^{[\alpha]}_{\tau,N-1} } ,
\end{equation}
where
\begin{equation}
d_{\tau,\alpha} \equiv 2\epsilon_{\alpha \alpha} + G_{\alpha \alpha } + 2(N-1)^2 \sum_{\beta \in O_\tau}^{\beta \neq \alpha}
A_{\alpha \beta}  (v_{\tau,\beta})^2 \frac{ \chi^{[\alpha \beta]}_{\tau,N-2}   }{   \chi^{[\alpha ]}_{\tau,N-1}   } , 
\end{equation}
\begin{equation}
\begin{aligned}
& \bra{\phi^{[\gamma]}_{\tau,N-1}} \hat{H} \ket{\phi^{[\gamma]}_{\tau,N-1}}   \equiv 
 \sum_{\alpha \in O_\tau}^{ \alpha \neq \gamma} (2 \epsilon_{\alpha \alpha}+G_{\alpha \alpha})  
\left( 1-\frac{\chi^{[\alpha\gamma]}_{\tau,N}}{\chi^{[\gamma]}_{\tau,N}} \right)\\
& \qquad\qquad\qquad\quad  + \sum_{\alpha \beta \in O_\tau }^{\alpha \neq \beta,  \alpha \neq \gamma, \beta \neq \gamma}  
\left[ G_{\alpha \beta }  N^2  v_{\tau,\alpha}    v_{\tau,\beta}  \frac{\chi^{[\alpha \beta\gamma]}_{\tau,N-1}}{\chi^{[\gamma]}_{\tau,N}} \right. \\
& \qquad\qquad\qquad\quad  \left. + A_{\alpha \beta}  \left( 1 - \frac {\chi^{[\alpha\gamma]}_{\tau,N} + \chi^{[\beta\gamma]}_{\tau,N} - \chi^{[\alpha \beta\gamma]}_{\tau,N}}  {\chi^{[\gamma]}_{\tau,N}}  \right) \right]. 
\end{aligned}
\end{equation}

\subsection{For a system with both valence protons and neutrons }

In this work, we extend the NBCS to the case of open-shell nuclides, that is, a pair-condensate ansatz with both valence protons and valence neutrons:
\begin{equation} \label{phi}
	\ket{\Phi} = \ket{\phi_{\pi,N_{\pi}}\phi_{\nu,N_{\nu}}} = \frac{1}{\sqrt{\chi_{\pi, N_\pi}\chi_{\nu, N_\nu}}  } (\hat{P}_{\pi}^\dagger)^{n_\pi} (\hat{P}_{\nu}^\dagger)^{n_\nu} \ket{0}.
\end{equation}
The Hamiltonian for open-shell nuclides is written as
 \begin{equation}
 \hat{H} = \sum_{\tau=\pi,\nu} ( \hat{H}_{0 \tau}+\hat{H}_{\tau\tau} ) + \hat{H}_{\pi \nu },
\end{equation}
where 
 \begin{equation}
\hat{H}_{\pi \nu } =  \sum_{\alpha \beta  \gamma \delta}  V_{\alpha \beta  \gamma \delta} \hat{a}^\dagger_{\pi,\alpha }  \hat{a}^\dagger_{\nu,\beta } \hat{a}_{\nu,\delta }  \hat{a}_{\pi,\gamma }.
\end{equation}
The energy is
\begin{equation}\label{totalenergy}
E= \sum_{\tau=\pi,\nu} E_{\tau} + E_{\pi \nu}, 
\end{equation}
where the valence proton energy $E_{\pi}$ and the valence neutron energy $E_{\nu}$ are obtained by Eq. (\ref{E1}), and the valence proton-neutron energy $E_{\pi \nu}$ is given by
\begin{equation}
\begin{aligned}
E_{\pi \nu}   \equiv & \bra{\Phi} \hat{H}_{\pi\nu} \ket{\Phi}\\
=& \sum_{\alpha \in O_\pi ,\beta \in O_\nu} 2A_{\alpha \beta }  
 \left( 1-\frac{\chi^{[\alpha]}_{\pi,N_\pi}}{\chi_{\pi,N_\pi}} \right) \left( 1-\frac{\chi^{[\beta]}_{\nu,N_\nu}}{\chi_{\nu,N_\nu}}  \right).
\end{aligned}
\end{equation}

The energy of an open-shell nucleus, $E$, varies with the pair structure coefficients $v_{\pi,\alpha}$ and $v_{\nu,\alpha}$. 
We determine $v_{\pi,\alpha}$ and $v_{\nu,\alpha}$ by minimizing $E$, which leads to the iterative formula
\begin{equation}\label{2.a.3} \small
	v_{\tau,\alpha}=\frac{\displaystyle -\sum_{\beta \in O_\tau}^{\beta \neq \alpha} G_{\alpha \beta } v_{\tau , \beta} \chi^{[\alpha \beta]}_{\tau,N_\tau-1} }{    (d_{\tau,\alpha}+ \bra{\phi^{[\alpha]}_{\tau,N_\tau-1}} H \ket{\phi^{[\alpha]}_{\tau,N_\tau-1}}-\bar{E}_\tau)    \chi^{[\alpha]}_{\tau,N_\tau-1} + Y^{[\alpha]}_{\tau,\gamma \beta}}.
\end{equation} 
Eq. (\ref{2.a.3}) differs from Eq. (\ref{2.a.2}) by the $Y^{[\alpha]}_{\tau,\gamma \beta}$ term, which itself is obtained from the variational principle for the proton-neutron energy $E_{\pi \nu}$, i.e.,
\begin{equation} \small
\begin{aligned}
Y^{[\alpha]}_{\tau,\gamma \beta} & =\sum_{\gamma \in O_\tau, \beta \in O_{{\tau}'}} 2 A_{\gamma \beta} \\
& \quad \times \frac{\chi^{[\alpha]}_{\tau,N_\tau-1}    \chi^{[\gamma]}_{\tau,N_\tau}   - \chi^{[\alpha \gamma]}_{\tau,N_\tau-1}    \chi_{\tau,N_\tau}  } {\chi_{\tau,N_\tau} }    \left( 1-\frac{\chi^{[\beta]}_{  {\tau}' ,N_{  {\tau}'}}}{\chi_{ {\tau}',N_{  {\tau}'}}} \right),
	\end{aligned}
\end{equation}
where ${\tau}'=\nu$ if $\tau=\pi$, and vice versa.

In our NBCS calculations, we use the conjugate gradient method to quickly minimize the energy $E$ [see Eq. (\ref{totalenergy})], and then use Eq. (\ref{2.a.3}) to iterate $v_{\pi,\alpha}$ and $v_{\nu,\alpha}$ until convergence.
Unsurprisingly, the NBCS energy is lower than the HF energy, as the ansatz wavefunction of a pair condensate
is more general than that of a Slater determinant.

\subsection{PNBCS: The angular momentum projection for NBCS}

In general, both the HF and NBCS break rotational symmetry. 
To cure this problem, projection on angular momentum is necessary.
While it is usually implemented as a numerical quadrature on Euler angles \cite{Edmonds1957,peterring}, 
a recently proposed linear algebra projection (LAP) ~\cite{PHF1} 
 is more than 10 times faster than the quadrature in nuclear structure computations \cite{PHF2}. 
 LAP has been successfully implemented to project nuclear states out of both Slater determinants~\cite{PHF1,PHF2} and out of a general pair condensate \cite{lu2022nuclear}.
Here we  outline the application of LAP to NBCS.

The $N$-pair condensate $\ket{\Phi}$ [see Eq. (\ref{phi})]  usually breaks rotational invariance. 
The condensate can be decomposed as a linear combination of normalized spherical tensors,
\begin{equation}
\ket{\Phi} = \sum_{JK} c_{JK} |J,K\rangle.
\end{equation}
The projection operator $\hat{P}^J_{MK}$ picks out the component $|J,K\rangle$, and rotates it to $|J, K\rightarrow M \rangle$, i.e.,
\begin{equation}
	\hat{P}^J_{MK} | \Phi \rangle = c_{JK} | J, K \rightarrow M \rangle.
\end{equation}
For states with angular momentum $J$, we diagonalize the Hamiltonian in the space spanned by 
\begin{equation}
	\left\{ \hat{P}^J_{M,-J} |\Phi \rangle, ~ \hat{P}^J_{M,-J+1} | \Phi \rangle,~ \cdots,~ \hat{P}^J_{M,J} | \Phi \rangle \right\},
\end{equation}
that is, we solve the discrete Hill-Wheeler equation:
\begin{equation}
	\sum_K {\mathcal H}^J_{K'K} g^r_{JK} = \epsilon_{J_r} \sum_K {\mathcal N}^J_{K' K} g^r_{JK}, \label{eqn:HillWheeler}
\end{equation}
where $\epsilon_{J_r}$ is energy of the $r$-th state with angular momentum $J$; 
$g^r_{JK}$ is the expansion coefficient of the projected eigenstate; 
and ${\mathcal H}^J_{K'K}$ and ${\mathcal N}^J_{K' K}$ are the Hamiltonian and norm matrix elements (or kernels), respectively:
\begin{equation}\label{2.c.2} \small
	\begin{aligned}
		{\mathcal H}^J_{K'K} =& \langle \hat{P}^J_{M K'} \Phi | \hat{H} | \hat{P}^J_{M K} \Phi \rangle
		= \langle \Phi | \hat{H} | \hat{P}^J_{K' K} \Phi \rangle
		, \\
		{\mathcal N}^J_{K'K} =&\langle \hat{P}^J_{M K'} \Phi |  \hat{P}^J_{MK} \Phi \rangle 
		= \langle \Phi |  \hat{P}^J_{K'K} \Phi \rangle
		.
	\end{aligned}
\end{equation}

In  traditional angular momentum projection, Eq. (\ref{2.c.2}) is calculated by numerical integrals:
\begin{equation} 
	\begin{aligned}
		{\mathcal H}^J_{K'K} =& \frac{2J+1}{8\pi^2}\int {\rm d}\Omega D^{J*}_{K'K}(\Omega) \langle \Phi | \hat{H} \hat{R}(\Omega) | \Phi \rangle
		, \\
		{\mathcal N}^J_{K'K} =&\frac{2J+1}{8\pi^2}\int {\rm d}\Omega D^{J*}_{K'K}(\Omega) \langle \Phi | \hat{R}(\Omega) | \Phi \rangle
		,
	\end{aligned}
\end{equation} 
where $D^J_{K'K}(\Omega)$ is the Wigner $D$ matrix.
In the LAP \cite{PHF1,PHF2}, the Hamiltonian and norm matrix elements are found by solving linear equations.
It is noticed that 
\begin{equation} \label{ad1}
	\begin{aligned}
		&\langle \Phi  |\hat{H}  \hat{R}(\Omega) | \Phi  \rangle \\
		&=  \sum_{JJ'KK'M  } c^*_{J'K'} c_{JK} D^J_{MK} (\Omega) \langle J' ,K' |\hat{H} | J ,K\rightarrow M \rangle \\
		&= \sum_{JKK'} D^J_{K'K}(\Omega) \mathcal{H}^J_{K'K},
	\end{aligned}
\end{equation}
and similarly 
\begin{equation} \label{ad2}
	\begin{aligned}
		&\langle \Phi  | \hat{R}(\Omega) | \Phi  \rangle = \sum_{JKK'} D^J_{K'K}(\Omega) \mathcal{N}^J_{K'K},
	\end{aligned}
\end{equation}
that is to say, $\langle \Phi  | \hat{H}\hat{R}(\Omega) | \Phi \rangle $ is a linear combination of $\mathcal{H}^J_{K'K}$, and $\langle \Phi  | \hat{R}(\Omega) | \Phi  \rangle $ is a linear combination of $\mathcal{N}^J_{K'K}$.
With a given set of Euler angles, one computes $\langle \Phi  | \hat{H}\hat{R}(\Omega) | \Phi \rangle $ and $\langle \Phi  | \hat{R}(\Omega) | \Phi  \rangle $, and then finds ${\mathcal H}^J_{K'K}$ and ${\mathcal N}^J_{K' K}$ as solutions of the linear Eqs. (\ref{ad1}) and (\ref{ad2}).

To compute $\langle \Phi  | \hat{R}(\Omega) | \Phi  \rangle $, we  use the Wigner $D$ matrix for SM single-particle states, i.e.,
\begin{equation}
\begin{aligned}
\mathcal{D}_{ab}(\Omega) 
&= \langle n_a l_a j_a m_a | \hat{R} (\Omega) | n_b l_b j_b m_b \rangle \\
&= \delta_{n_a n_b } \delta_{ l_a l_b } \delta_{ j_a j_b } D^{j_a}_{m_a m_b }(\Omega), 
\end{aligned}
\end{equation}
and rewrite a collective pair [see Eq. (\ref{3})] as
\begin{equation}
\begin{aligned}
\hat{P}_{\tau}^\dagger &= \sum_{\alpha \in O_\tau} v_{\tau,\alpha} \hat{a}^\dagger_{\tau,\alpha} \hat{a}^\dagger_{\tau, \tilde{\alpha}}\\
&=  -\sum_{ab;\alpha \in O_\tau} v_{\tau,\alpha}  U_{\alpha a} U_{\alpha \tilde{b}} \hat{c}^\dagger_{\tau,a} \hat{c}^\dagger_{\tau, b} \\
&= \sum_{ab} p_{ab} \hat{c}^\dagger_{\tau,a} \hat{c}^\dagger_{\tau, b}.
\end{aligned}
\end{equation}
Under rotational transformation, the collective pair becomes
\begin{equation}
\hat{R}(\Omega) \hat{P}_{\tau}^\dagger \hat{R}^{-1}(\Omega) = \hat{P}_{\tau}^{'\dagger}
\end{equation}
with $p' =  \mathcal{D} p \mathcal{D}^\top $.
The matrix elements of $\hat{R}(\Omega)$ for the NBCS is
\begin{equation}\label{2.c.3}
\bra{ (\hat{P}_{\tau}^\dagger)^N}  \hat{R}(\Omega)  \ket{(\hat{P}_{\tau}^\dagger)^N} = \bra{ (\hat{P}_{\tau}^\dagger)^N} \ket{(\hat{P}_{\tau}^{'\dagger})^N } ,
\end{equation} 
and similarly
\begin{equation}\label{2.c.4}
\bra{ (\hat{P}_{\tau}^\dagger)^N} \hat{H} \hat{R}(\Omega)  \ket{(\hat{P}_{\tau}^\dagger)^N} = \bra{ (\hat{P}_{\tau}^\dagger)^N}\hat{H} \ket{(\hat{P}_{\tau}^{'\dagger})^N } .
\end{equation} 
A fast algorithm for computing Eqs. (\ref{2.c.3}) and (\ref{2.c.4}) was given in Ref. \cite{lu2022nuclear}.

\begin{figure*}
	\centering
	\includegraphics[width=0.7\linewidth]{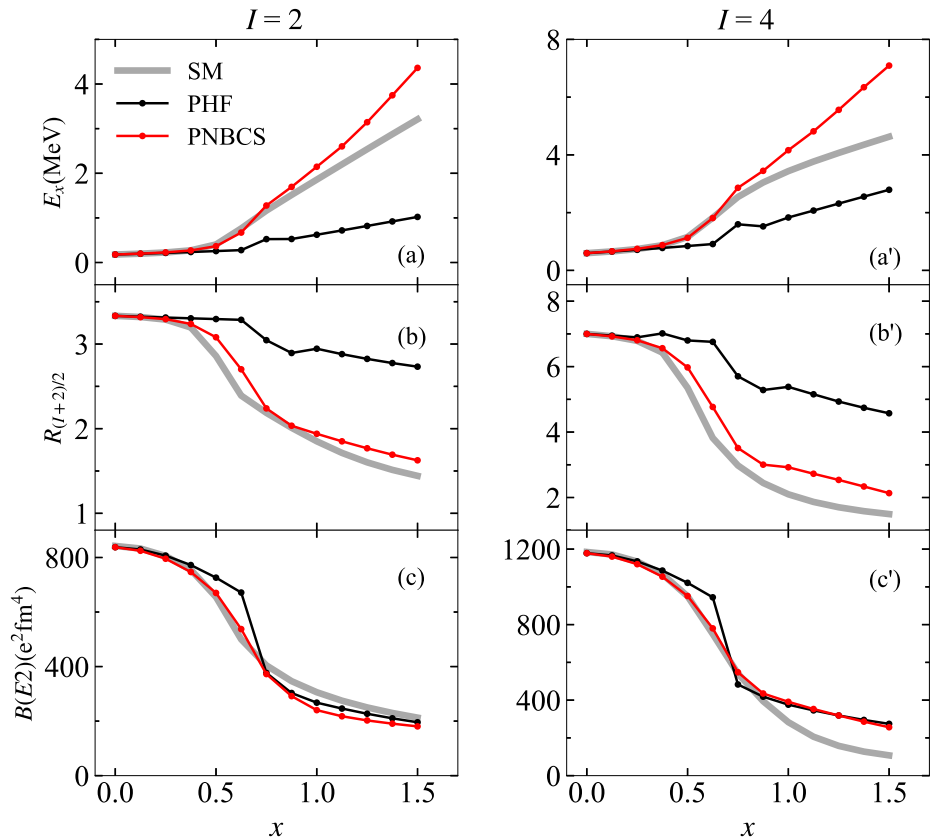}
	\caption{The excitation energy $E_x(I^+)$, the level energy ratios $R_{(I+2)/2}$ (where $R_{I/2}\equiv E_x(I^+)/E_x(2^+)$), and the $B(E2;I \rightarrow I-2)$ values for 6 protons and 6 neutrons in the $pf$ shell with the schematic interaction $\hat{H}(x)$ [see Eq. (\ref{3.1})]. Panels (a), (b), (c) correspond to $I = 2$ and Panels (a$^\prime$), (b$^\prime$), (c$^\prime$) to $I = 4$. SM is the abbreviation for the full shell model; PHF is for the angular momentum projected Hartree-Fock; and PNBCS is for the angular momentum projected particle number conserved BCS. 
	}
	\label{fig:check-pairing}
\end{figure*}

\section{Parameter-driven shape evolution in PNBCS }

We investigate the validity and utility of the PNBCS as the nuclear shape evolves from spherical to quadrupole deformation. 
Here the shape evolution is realized by changing the ratio of pairing and quadrupole-quadrupole interactions in a schematic Hamiltonian
\begin{equation}\label{3.1}
	\hat{H}(x)=x \left(  \sum_{j_a}   \epsilon_{j_a}  \hat{n}_{j_a}  + g \hat{V}_{\rm P} \right)   +  \kappa \hat{V}_Q
\end{equation} 
The first term $\sum_{j_a} \epsilon_{j_a} \hat{n}_{j_a}$ is single particle energy term.
The second term is the monopole pairing interaction:
\begin{equation}
	\hat{V}_{\rm P} = - \hat{A}^{(0)\dagger}_{\pi} \hat{A}^{(0)}_{\pi} -  \hat{A}^{(0)\dagger}_{\nu} \hat{A}^{(0)}_{\nu} ,
\end{equation} 
\begin{equation}
	{ \hat{A}^{(0)\dagger} } = \displaystyle \sum_{j_{a}} \frac{\sqrt{2j_{a}+1}}{2} \left( \hat{c}_{j_{a}}^{\dagger} \times \hat{c}_{j_{a}}^{\dagger} \right)^{(0)} .
\end{equation} 
The last term is the quadrupole-quadrupole interaction:
  \begin{equation}
\hat{V}_Q = -(\hat{Q}_\pi+\hat{Q}_\nu) \cdot (\hat{Q}_\pi+\hat{Q}_\nu).
\end{equation}

We study the system of six protons and six neutrons in the $pf$ shell using the schematic Hamiltonian.
The single particle energies are taken from the KB3G effective interaction \cite{POVES2001157}, i.e, 
$\epsilon_{0f_{7/2}}=0$ MeV, 
$\epsilon_{1p_{3/2}}=2.0$ MeV, 
$\epsilon_{0f_{5/2}}=6.5$ MeV, 
$\epsilon_{0p_{1/2}}=4.0$ MeV.
The strength parameter of the monopole pairing interaction and quadrupole-quadrupole interaction are set to be $g=0.4$ Mev and $\kappa=0.1$ MeV, respectively.
The adjustable parameter $x$ ranges from 0 to 1.5.

We calculate level energies and the electric quadrupole reduced
transition probabilities $B(E2)$ (taking the standard effective charges $e_\pi$=1.5 and  $e_\nu$=0.5) in the SM using the {\sc Bigstick}  code \cite{bigstick,johnson2018bigstick}, and in  angular momentum projected HF (PHF) and in  PNBCS using unpublished codes.
Fig. \ref{fig:check-pairing} shows the results of the excitation energies $E_x(I^+_1)$, the energy ratios $R_{(I+2)/2}$ (where $R_{I/2} \equiv E_x(I^+_1) / E_x(2^+_1)$), and the $B(E2;I \rightarrow I-2)$ values with $I=2$ and 4.
The SM result exhibits nuclear shape evolution.
For small $x$, i.e., dominated by the quadrupole-quadrupole interaction, the $2_1^+$ and $4_1^+$ excitation energies are small with  the energy ratios $R_{4/2} \sim 3.33$ and $R_{6/2} \sim 7 $, and the $B(E2)$ values are large.
These are typical features of rotational nuclei in the Elliott's SU(3) dynamical symmetry limit.
As $x$ increases, the $2_1^+$ and $4_1^+$ excitation energies increase and the $B(E2)$ values decrease. 
For large $x$, i.e., large monopole pairing interaction and single-particle splittings, the $2_1^+$ and $4_1^+$ excitation energies are large and the energy ratios and the $B(E2)$ values are small, typical behavior of spherical nuclei.

In general the PNBCS result is better that the PHF one.
In the limiting case of $x \sim 0$, that is, large quadrupole-quadrupole interaction, both the PHF and PNBCS well reproduce the excitation energies, energy ratios, and $B(E2)$ values.
This result can be understood as follows.
The pure quadrupole-quadrupole interaction in a harmonic-oscillator shell (i.e., Elliott's model) provides us with a HF state with an axially symmetric quadrupole deformed shape. 
From this HF state, one can project out the exact ground rotational band.
For $ 0.5 < x <0.75 $, the PNBCS also provides a good description for the $2_1^+$ and $4_1^+$ states of transitional nuclei with $ 2.2 < R_{4/2} < 2.8$, while the PHF result quickly deteriorates.
In the limiting case of $x \sim 1.5$, that is, large monopole pairing interaction and single-particle splittings, neither the PHF nor PNBCS is good, and it is expected that the PNBCS generalized with broken-pair configurations, that is, 
a multi-reference PNBCS related to generator-coordinate and other beyond-mean-field methods, will greatly improve the results, which we leave to future work.

\section{For nuclei in the medium-heavy mass region}

In this section, we show that the PNBCS provides us reasonably good description for low-lying states of transitional and deformed nuclei.
We exemplify this with $^{44,46,48}$Ti, $^{48,50}$Cr, $^{52}$Fe in the $pf$ shell, $^{60,62,64}$Zn, $^{66,68}$Ge, $^{68}$Se in the $pf_{5/2}g_{9/2}$ space, and $^{108,110}$Xe in the $sdg_{7/2}h_{11/2}$ space, by comparing the results of the SM, PHF, and PNBCS.
We also predict low-lying levels for $^{112,114}$Ba and $^{116,118,120}$Ce.

\begin{figure}
	\centering
	\includegraphics[width=1\linewidth]{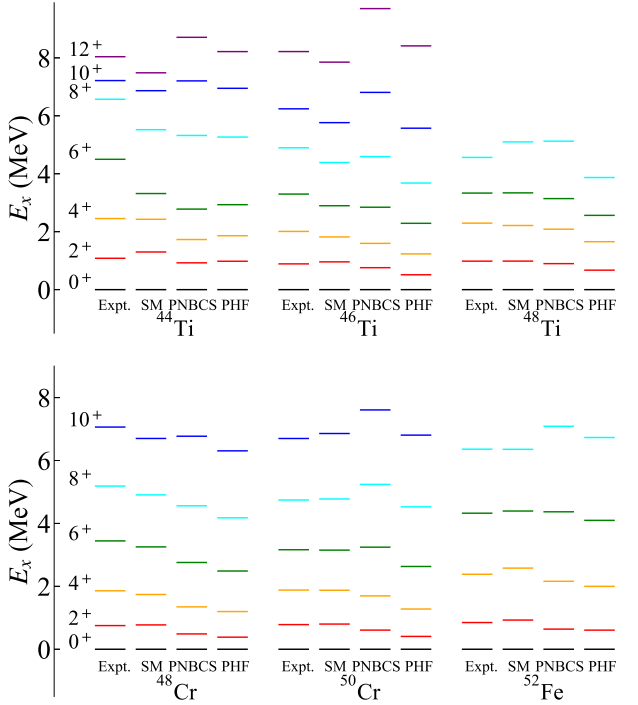}
	\caption{The yrast states of $^{44,46,48}\textrm{Ti}$, $^{48,50}\textrm{Cr}$, and $^{52}\textrm{Fe}$. 
		SM is the abbreviation for the full shell model; PNBCS is for the angular momentum projected particle number conserved BCS; and PHF is for the angular momentum projected Hartree-Fock. The experimental data are taken from Refs. \cite{WU20001,BURROWS20061747,CHEN20112357,CHEN20191,DONG2015185}.
	}
	\label{fig:pf}
\end{figure}

\begin{table}\center
	\caption{\label{table1} 
		$B(E2; I \rightarrow I-2)$ (in W.u.) for the yrast states of $^{44,46,48}\textrm{Ti}$, $^{48,50}\textrm{Cr}$, and $^{52}\textrm{Fe}$. The experimental data are taken from Refs. \cite{WU20001,BURROWS20061747,CHEN20112357,CHEN20191,DONG2015185}.
	}
	\begin{tabular}{cc    cc    cc    cc   cc   ccc}  \hline\hline
		& Nuclide  &&  $I^{\pi}$  && Expt. && SM &&  PNBCS && PHF  &   \\  \hline
		& \multirow{4}{*}{$^{44}$Ti}     && $2^+$ && 13(4)    && 12.9 && 12.9&&13.7&\\
		&                                                            && $4^+$ && 30(5)    && 17.0 && 18.3&&18.9  &\\
		&                                                            && $6^+$ && 17.0(24) && 14.2 && 18.3&&18.8  &\\
		&                                                            && $8^+$ &&   -      && 10.1 && 16.5&&16.9 &\\
		\hline
		& \multirow{4}{*}{$^{46}$Ti}     && $2^+$ && 19.5(6)   && 12.9 && 14.2&&15.1&\\
		&                                                            && $4^+$ && 20.2(13)  && 17.0&& 20.6 &&21.0 &\\
		&                                                           && $6^+$ && 16.4(15)  && 17.2&& 21.6 &&21.7 &\\
		&                                                            && $8^+$ && 11.3(14)  && 15.7 && 20.0&&20.1&\\
		\hline
		& \multirow{4}{*}{$^{48}$Ti}            && $2^+$ && 14.7(4)  && 10.1 && 11.2 &&11.6&\\
		&                                                            && $4^+$ && 18.4(17) && 15.0 && 16.7 &&16.5 &\\
		&                                                            && $6^+$ &&    -     && 6.2  && 17.0&&16.5  &\\
		&                                                            && $8^+$ &&    -     && 5.1  && 14.8  &&13.9&\\
		\hline
		& \multirow{4}{*}{$^{48}$Cr}           && $2^+$ && 31(4) && 20.6 && 24.0 &&24.6&\\
		&                                                            && $4^+$ && 27(3) && 28.2 && 34.2&&34.4&\\
		&                                                            && $6^+$ && 29(8) && 28.3 && 36.9&&36.5&\\
		&                                                            && $8^+$ && 24(7) && 26.2 && 36.9 &&36.0&\\
		\hline
		& \multirow{4}{*}{$^{50}$Cr}     && $2^+$ && 19.3(6) && 16.9 && 19.7&&20.2 &\\
		&                                                            && $4^+$ && 14.6(16) && 24.0 && 28.6 &&28.1&\\
		&                                                            && $6^+$ && 22(5) && 20.4 && 31.0&&29.3&\\
		&                                                            && $8^+$ && 19(5) && 17.6 && 30.7 &&28.4&\\
		\hline
		& \multirow{4}{*}{$^{52}$Fe}     && $2^+$ && 14.2(19) && 16.0 && 15.8 &&15.7&\\
		&                                                            && $4^+$ && 26(6) && 21.3 && 22.1 &&21.5&\\
		&                                                            && $6^+$ && 10(3) && 11.8 && 22.8 &&22.0&\\
		&                                                            && $8^+$ && 9(4) && 7.2 && 20.9 &&20.4&\\
		\hline
		\hline\hline
	\end{tabular}
\end{table}
	
\subsection{$^{44,46,48}$Ti, $^{48,50}$Cr, $^{52}$Fe in the $pf$ shell}

$^{44,46,48}\textrm{Ti}$, $^{48,50}\textrm{Cr}$, and $^{52}\textrm{Fe}$ are typical transitional and deformed nuclei in the $pf$ shell.
In our calculation, we use the KB3G interaction \cite{POVES2001157} and take the standard effective charges $e_\pi =1.5$ and $e_\nu = 0.5$ for $B(E2)$s.
The low-lying states are well reproduced by the SM calculation, except for $^{44}\textrm{Ti}$.

Fig. \ref{fig:pf} and Table \ref{table1} compare the excitation energies and $B(E2;I \rightarrow I-2)$ values for the yrast states of $^{44,46,48}\textrm{Ti}$, $^{48,50}\textrm{Cr}$, and $^{52}\textrm{Fe}$ from the experimental data \cite{WU20001,BURROWS20061747,CHEN20112357,CHEN20191,DONG2015185}, the full SM, the PNBCS, and the PHF.
The PNBCS results are in good agreement with the data and/or the SM results.
The level energies of the $2^+$, $4^+$, $6^+$, $8^+$ states obtained by the PNBCS are very close to those by the SM.

The level energies obtained by the PHF are visibly smaller than those by the PNBCS,
and the $B(E2)$ values obtained by the former are slightly larger than those by the latter.
This indicates that the pair correlation plays an important role in reduction of the moment of inertia but plays a limited one in reduction of the electric quadrupole transition probability;
the electric quadrupole transition strength is affected mainly by the deformation encoded in HF.
The PNBCS results may be improved if the variation of the HF single particle states is simultaneously taken into account within the NBCS variation, and result will be equivalent to that of the particle number projected Hartree-Fock-Bogoliubov \cite{peterring,Dietrich1964}, which we leave to future work.

\begin{figure}
	\centering
	\includegraphics[width=1\linewidth]{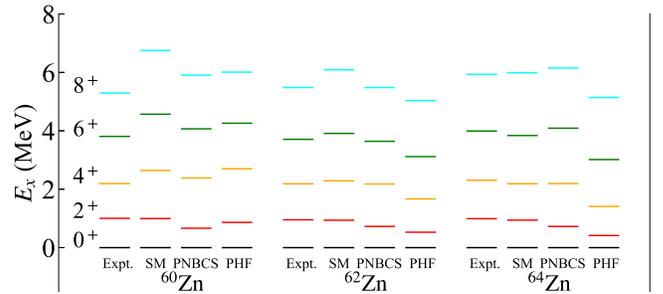}
	\caption{The yrast state of $^{60,62,64}\textrm{Zn}$. The experimental data are taken from Refs.  \cite{BROWNE20131849,NICHOLS2012973,SINGH2007197}.}
	\label{fig:zn}
\end{figure}

\begin{table}\center
	\caption{\label{table2} $B(E2; I \rightarrow I-2)$ (in W.u.) for the yrast states of $^{60,62,64}\textrm{Zn}$. The experimental data are taken from refs. \cite{NICHOLS2012973,PRITYCHENKO2012798,SINGH2007197}.
	}
	\begin{tabular}{cc    cc    cc    cc  cc  ccc}  \hline\hline 
		& Nuclide  &&  $I^{\pi}$  && Expt. && SM &&  PNBCS&&PHF  &   \\  \hline
		& \multirow{4}{*}{$^{60}$Zn}          && $2^+$ &&   - && 10.9&& 10.3  &&10.6&\\
		&                                				                  && $4^+$ &&  -  && 12.8  &&13.0 &&13.0&\\
		&                                			  	                 && $6^+$ &&  - &&11.9  && 10.8  &&10.4&\\
		&                               				                 && $8^+$ &&  - && 7.4  && 6.1  &&5.7&\\
		\hline
		& \multirow{4}{*}{$^{62}$Zn}           && $2^+$ && 16.8(8)              && 11.1 &&  11.5    &&12.1&\\
		&                              					                  && $4^+$ && 26($^{+7} _{-12}$)      && 10.0 &&  12.5  &&12.3&\\
		&                              					                  && $6^+$ && 19(3)                 && 14.2 &&  12.3  &&14.2&\\
		&                                				                  && $8^+$ && 7.9($^{+20} _{-40}$)  && 12.1  &&  10.4 &&12.2 &\\
		\hline
		& \multirow{4}{*}{$^{64}$Zn}          && $2^+$ && 20.0(6)  && 10.6  && 11.7    &&12.4 &\\
		&                                   			           	  && $4^+$ && 12.2(5)  && 12.7  && 14.9   &&16.4&\\
		&                                   			           	  && $6^+$ && 23(6)    && 13.3  && 14.2   &&15.4 &\\
		&                                     			           	  && $8^+$ &&   -         && 14.8  && 11.6   &&12.1 &\\
		\hline
		
		\hline\hline
	\end{tabular}
\end{table} 

\begin{figure}
	\centering
	\includegraphics[width=1\linewidth]{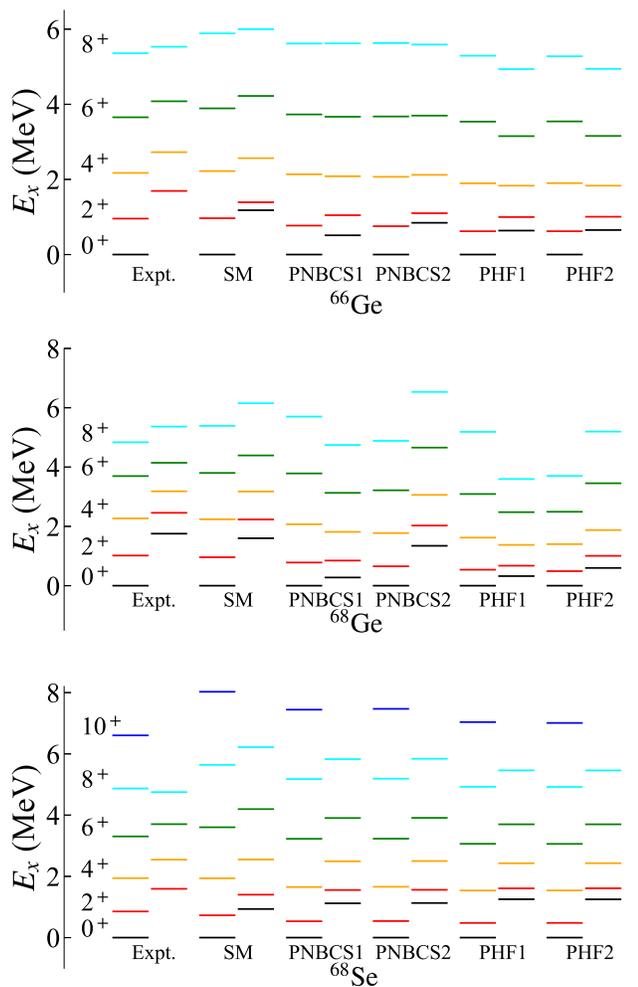}
	\caption{The ground rotational band and the side band of $^{66}\textrm{Ge}$, $^{68}\textrm{Ge}$, and $^{68}\textrm{Se}$. The experimental data are taken from Refs. \cite{BROWNE20101093,MCCUTCHAN20121735}. }
	\label{fig:ge}
\end{figure}

\begin{table*}\center
	\caption{\label{table3} $B(E2; I \rightarrow I-2)$ (in W.u.) for the ground and side bands of $^{66}\textrm{Ge}$, $^{68}\textrm{Ge}$, and $^{68}\textrm{Se}$.  The experimental data are taken from Refs. \cite{BROWNE20101093,MCCUTCHAN20121735}.
	}
	\begin{tabular}{cc    cc    cc    cc    cc  cc cc ccc}  \hline\hline 
		& Nuclide  &&  $I^{\pi}$  && Expt. && SM &&  PNBCS1&&   PNBCS2&&PHF1&&PHF2 &   \\  \hline
		& \multirow{4}{*}{$^{66}$Ge (ground band)}  && $2^+$ && 12.0(23)        && 15.9 &&  12.3 && 12.7  &&12.4&&12.4&\\
		&                                    							   			&& $4^+$ && $>9.6$         && 18.2 &&  16.7&&  14.3&& 16.5&&13.8&\\
		&                                     							  			&& $6^+$ && $>1.2$          && 19.9 &&  16.8 && 17.4&&15.8&&14.4&\\
		&                                     							  			&&$8^+$ && -                     && 9.0&& 14.9&& 16.1 &&13.2&&12.8 &\\
		\hline
		& \multirow{4}{*}{$^{66}$Ge (side band)}         && $2^+$ && -    &&6.7 &&  17.3 &&  12.4&&18.9 &&18.7&\\
		&                                   						 	 			&& $4^+$ && -      &&11.4&&  22.4&&   11.8 &&25.0 &&21.7&\\
		&                                    										&& $6^+$ &&-       &&13.0 &&  23.2 && 10.4 &&24.0 &&22.2&\\
		&                                   							 			&& $8^+$ && -     &&1.6&&  14.8&& 11.2  &&16.9&&16.9&\\
		\hline
		& \multirow{4}{*}{$^{68}$Ge (ground band)} && $2^+$  && 15.3(8)               && 15.4&&  13.2&&   14.2    &&11.8 &&14.5&\\
		&                                    							   && $4^+$ &&12.8(15)       && 21.3&&  16.4&&      20.2&& 16.2&&21.4&\\
		&                                     							 	&& $6^+$ && 12 (4)         && 24.1&&  17.6 &&      27.2&&13.2&&31.9&\\
		&                                     							  &&$8^+$ && 14 (3)             && 16.6&&  9.8 &&      28.6 &&8.8 &&37.5&\\
		\hline
		& \multirow{4}{*}{$^{68}$Ge (side band)}         &&$2^+$ && -    &&10.7&& 21.2 &&14.7      &&24.4&&18.5   &\\
		&                                   						 				&& $4^+$ && -  && 12.7&&  27.6&&  16.6   &&32.2&&  10.0&\\
		&                                    										 && $6^+$ &&-  && 8.3&&   31.6&&  13.0   &&37.9&&   10.6&\\
		&                                    										 && $8^+$ &&-  && 5.0 &&   31.1&&  9.5   && 38.7 &&  8.6  &\\
		\hline
		& \multirow{4}{*}{$^{68}$Se (ground band)}                && $2^+$ && 27(4)       && 21.1&&  22.4&& 22.4  &&23.0&& 23.0 &\\
		&                                    							   							&& $4^+$ &&-           && 30.1&&30.5&& 30.5 &&31.0&& 31.0 &\\
		&                                     							 							 && $6^+$ && -         && 30.8&&  30.5&& 30.5 &&30.4&&  30.4&\\
		&                                     							  							&&$8^+$ && -           && 23.7&&  26.9 &&  26.9   &&26.4&&  26.4 &\\	
		\hline
		& \multirow{4}{*}{$^{68}$Se (side band)}                            && $2^+$ && -     &&19.4&&  20.1&& 20.1 &&21.0&& 21.0 &\\
		&                                   						 	 							&& $4^+$ && -      && 26.8&&  27.7&& 27.7 &&28.6&& 28.6 &\\
		&                                    							 							&& $6^+$ &&-       &&26.8&&  28.0&& 28.0 &&28.5&& 28.5 &\\
		&                                   							 							 && $8^+$ && -     &&22.7&&  24.9 && 24.9 &&25.1&& 25.1 &\\
		\hline\hline
	\end{tabular}
\end{table*} 

\subsection{$^{60,62,64}$Zn, $^{66,68}$Ge, $^{68}$Se in the $pf_{5/2}g_{9/2}$ space}

We calculate low-lying states of $^{60,62,64}$Zn, $^{66,68}$Ge, and $^{68}$Se in the $pf_{5/2}g_{9/2}$ space (i.e., the $1p_{1/2}1p_{3/2}0f_{5/2}0g_{9/2}$ space), using the JUN45 interaction \cite{PhysRevC.80.064323,HJORTHJENSEN1995125} and the standard effective charges $e_\pi =1.5$ and $e_\nu = 0.5$ for $B(E2)$s.
The deformations of these nuclei are not very strong.
Therefore the low-lying states are appropriately reproduced by the above SM calculations.

The $^{60,62,64}$Zn isotopes are transitional nuclei.
Fig. \ref{fig:zn} and Table \ref{table2} compare for the yrast states the experimental data \cite{BROWNE20131849,NICHOLS2012973,PRITYCHENKO2012798,SINGH2007197},  full SM,  PNBCS, and  PHF.
Both the level energies and the $B(E2)$ values obtained by the PNBCS are very close to the data or the SM results.
For $^{62}$Zn and $^{64}$Zn, the moment of inertia and the electric quadrupole transition probability obtained by  PHF is (slightly) larger than those by  PNBCS, similar to the cases in the $pf$ shell.

Shape coexistence \cite{Heyde-rmp} has been observed in $^{66,68}$Ge and $^{68}$Se \cite{BROWNE20101093,MCCUTCHAN20121735}, and both the ground and side rotational bands are well reproduced by the SM calculation (Fig. \ref{fig:ge}).

For $^{66}$Ge, our HF calculation produces an oblate minimum with $ \braket{\beta}=0.21$ and $ \braket{\gamma}=60^{\circ}$ and a triaxially deformed one with $ \braket{\beta}=0.24$ and $ \braket{\gamma}=8^{\circ}$, separated only by 0.58 MeV,
and the configuration spaces constructed by the angular momentum projection on them are denoted by $L_1$ and $L_2$, respectively.
We carry out the PHF calculation in two different ways:
(1) the oblate and $\gamma$ bands are calculated by solving the Hill-Wheeler equation in the $L_1$ and $L_2$ spaces, respectively (the calculation results are denoted by PHF1);
(2) the bands are calculated in the $L_1 \oplus L_2$ space, i.e., the configuration mixing between the two HF states is allowed (denoted by PHF2).
Similarly, we carry out the PNBCS calculation with and without the configuration mixing (denoted by PNBCS2 and PNBCS1), respectively.

In Fig. \ref{fig:ge} we see the energy levels in both the ground and side $\gamma$ bands of $^{66}$Ge are well reproduced by the calculations, including PNBCS1, PNBCS2, PHF1, and PHF2.
In Table \ref{table3} we see for the $B(E2)$ values in the ground band, the results of PNBCS1, PNBCS2, PHF1, and PHF2 are all in good agreement with the SM one,
but for the $B(E2)$ values in the side $\gamma$ band, only the PNBCS2 result are close to the SM one.
The above results indicate that both the pair correlation and the configuration mixing between the oblate and triaxially deformed states are important in reproducing the electric quadrupole transition rate in $^{66}$Ge.
Both the SM and the PNBCS2 predict the $\gamma$ band head is a $0^+$ state, which has not yet been found experimentally.

For $^{68}$Ge, our HF calculation with Kramers degeneracy produces two minima differing in energy by only 0.76 MeV, both of which are triaxially deformed: the first minimum has $\braket{\beta}=0.17$ and $\braket{\gamma}=38^{\circ}$, and the second one has $\braket{\beta}=0.24$ and $\braket{\gamma}=18^{\circ}$.
It is worth mentioning that the probability of the second minimum showing up in the HF calculation with Kramers degeneracy is $\sim 3 \%$, and that in the time-reversal-unconstrained HF calculation is $\sim 0.03 \%$.
Similar to the case of $^{66}$Ge, both the PNBCS and PHF calculations are carried out in two different ways, i.e., PNBCS1, PNBCS2, PHF1, and PHF2.
The results are presented in Fig. \ref{fig:ge} and Table \ref{table3}.
The PNBCS2 results are in good agreement with the data or the SM results, but the PNBCS1, PHF1, and PHF2 fail to reproduce the level energies of the side band.
For the $B(E2)$ values of the side band, the PNBCS2 and PHF2 results are much better than the PNBCS1 and PHF1 ones.
Both the pair correlation and the configuration mixing between the two HF states are important.

For $^{68}$Se, our HF calculation produces an oblate minimum with $ \braket{\beta}=0.22$ and $ \braket{\gamma}=60^{\circ}$ and a prolate one with $ \braket{\beta}=0.21$ and $ \braket{\gamma}=0^{\circ}$.
Our PNBCS1, PNBCS2, PHF1, and PHF2 results are all closed to the SM results. 
The pair correlation and the configuration mixing between the oblate and prolate states are not important here.

\begin{figure}
	\centering
	\includegraphics[width=1\linewidth]{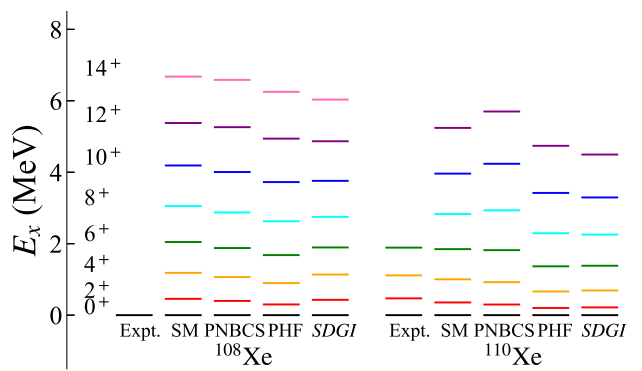}
	\caption{ The ground band of $^{108,110}$Xe. 
		$SDGI$ represents the result calculated by the $SDGI$-pair approximation in Ref. \cite{npadeformed3}.
		The experimental data are taken from Ref. \cite{GURDAL20121315}.
	}
	\label{fig:xe}
\end{figure}

\begin{table}\center
	\caption{\label{table4} $B(E2; I \rightarrow I-2)$ (in W.u.) for the ground band of $^{108,110}$Xe. 
	}
	\begin{tabular}{cc    cc    cc   cc cc  cc  ccc}  \hline\hline 
		& Nuclide  &&  $I^{\pi}$  && Expt. && SM &&  PNBCS&&PHF && $SDGI$ &   \\  \hline
		& \multirow{5}{*}{$^{108}$Xe}          && $2^+$ && - && 30.9 &&  30.6&&32.4&& 28.3&\\
		&                                                                 && $4^+$ && - && 41.8&&  39.9&&44.1&& 38.1&\\
		&                                                                 && $6^+$ && - && 44.9&& 45.4&&48.8&& 38.7&\\
		&                                                                 && $8^+$ && - && 47.4 &&  48.2&&50.9&& 41.6&\\
		&                                                                 && $10^+$ && - && 44.8 &&  47.0&&49.3&&40.2&\\
		&                                                                 && $12^+$ && - && 39.3 &&  43.5&&45.3&&36.0&\\
		\hline
		& \multirow{4}{*}{ $^{110}$Xe}                              && $2^+$ && -     && 34.3&&35.1&&37.0&& 36.2&\\
		&                                                                 					&& $4^+$ && -    && 48.4&&  49.5&&52.0&& 51.1&\\
		&                                     				                            	&& $6^+$ &&-       && 51.7  &&  53.2&&55.5&&54.8 &\\
		&                                    			                            		     && $8^+$ && - && 52.1 &&53.5&&55.4&& 54.5&\\
		&                                  				                            	       && $10^+$ && - && 50.8 &&  51.9&&53.2&&51.8&\\
		&                                    			                            		   && $12^+$ && - && 46.9&&  48.6&&49.4&&47.4&\\
		\hline
		
		\hline\hline
	\end{tabular}
\end{table}

\subsection{$^{108,110}$Xe, $^{112,114}$Ba, $^{116,118,120}$Ce in the $sdg_{7/2}h_{11/2}$ space}

The nuclides around the $N=Z$ line above $^{100}$Sn are of great interest due to their potential importance in the study of superallowed $\alpha$ decay and nucleosynthesis.
In this work, we calculate low-lying states of $^{108,110}$Xe, $^{112,114}$Ba, $^{116,118,120}$Ce in the $sdg_{7/2}h_{11/2}$ space (i.e., the $2s_{1/2}1d_{3/2}1d_{5/2}0g_{7/2}0h_{11/2}$ space), using the monopole-optimized effective interaction based on the CD-Bonn potential renormalized by the perturbative G-matrix approach \cite{private1}, and the standard effective charges $e_\pi =1.5$ and $e_\nu = 0.5$ for $B(E2)$s.

The $N=Z$ nuclide $^{108}$Xe has been observed recently {\cite{BLACHOT2000135}}, but the low-lying spectrum remains unknown.
The excitation energies of low-lying states of its neighbor $^{110}$Xe has been measured {\cite{GURDAL20121315}}, showing collective rotational features, and are well reproduced by the SM calculation (see Fig. \ref{fig:xe}).
Fig. \ref{fig:xe} and Table \ref{table4} compare for the ground band of $^{108,110}$Xe from the full SM, PNBCS, PHF (and the experimental data for $^{110}$Xe).
Since the nucleon-pair approximation \cite{npa3,npaodd,npareports} is an efficient truncation model of the full SM, the results obtained by the $SDGI$-pair approximation \cite{npadeformed3} are also included for comparison.
As shown in Fig. \ref{fig:xe} and Table \ref{table4}, the PNBCS results are in good agreement with the data or the SM results.
The excitation energies obtained by the PHF and the $SDGI$-pair approximation are slightly lower.

\begin{figure*}
	\centering
	\includegraphics[width=0.6\linewidth]{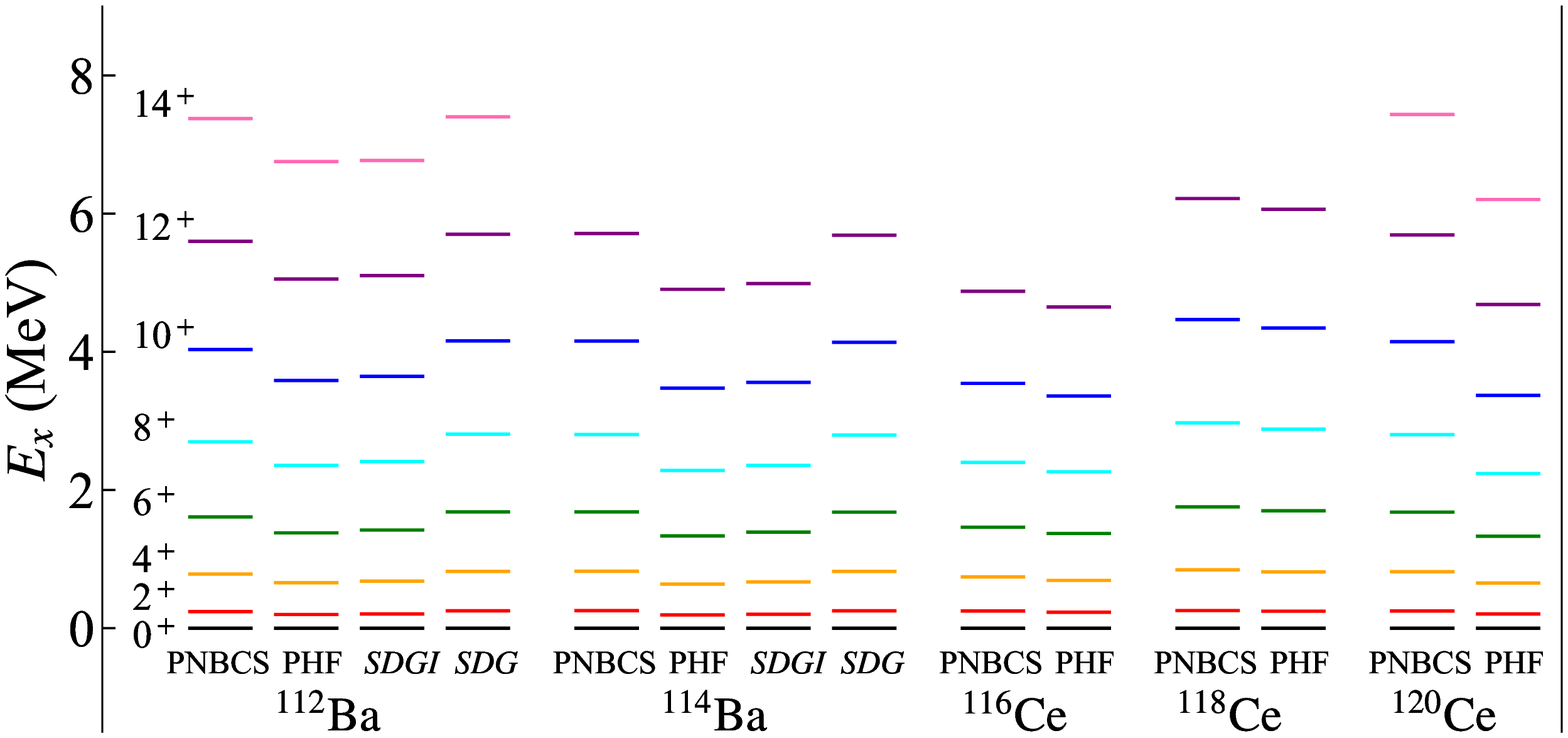}
	\caption{ The ground band of $^{112,114}$Ba and $^{116,118,120}$Ce. }
	\label{fig:ba}
\end{figure*}

\begin{table}\center
	\caption{\label{table5} $B(E2; I \rightarrow I-2)$ (in W.u.) for the ground band of $^{112,114}$Ba and $^{116,118,120}$Ce.
	}

	\begin{tabular}{cc   cc cc cc cc ccc}  \hline
		& Nuclide  &&  $I^{\pi}$   && PNBCS&&PHF&& $SDGI$ &&  $SDG$ &   \\  \hline
		& \multirow{5}{*}{$^{112}$Ba}         && $2^+$  &&52.2&&54.4&& 54.0  && 50.3  &\\
		&                                                                 && $4^+$ &&73.7&&76.5&&  76.0 && 70.7  &\\
		&                                                                && $6^+$ &&79.3&&81.9 && 81.3  &&  75.5 &\\
		&                                                                 && $8^+$ &&80.0&&82.2&&  81.5  && 75.6 &\\
		\hline
		& \multirow{5}{*}{$^{114}$Ba}                      && $2^+$  &&54.4&&55.6&& 54.6  && 50.2  &\\
		&                                                                		 && $4^+$ &&77.0&&78.2&& 76.9  && 70.6  &\\
		&                                                                 		&& $6^+$  &&82.9&&83.8&& 82.4  &&  75.5 &\\
		&                                                                		&& $8^+$  &&83.8&&84.2&& 82.8  && 75.9  &\\
		\hline
		& \multirow{5}{*}{$^{116}$Ce}                && $2^+$ &&74.4 &&76.7&& -  && -&\\
		&                                                                 && $4^+$  &&101.4 &&104.7&& -  && -&\\
		&                                                                 && $6^+$ &&125.3 &&129.2&& -  && -&\\
		&                                                                 && $8^+$ &&136.1 &&140.0&& -  && -&\\
		
		\hline
		& \multirow{5}{*}{$^{118}$Ce}                 && $2^+$  &&64.3&&  64.0 && -  && -&\\
		&                                                                 && $4^+$  &&91.1 &&  92.5 && -  && -&\\
		&                                                                 && $6^+$  &&98.6 && 97.9  && -  && -&\\
		&                                                                 && $8^+$  &&100.8&&   99.3&& -  && -&\\
		
		\hline
		& \multirow{5}{*}{$^{120}$Ce}               && $2^+$  &&67.9&& 79.9 && -  && -&\\
		&                                                                 && $4^+$  &&97.9&& 107.7&& -  && -&\\
		&                                                                 && $6^+$ &&109.6 && 131.8 && -  && -&\\
		&                                                                 && $8^+$ &&117.4&&  142.8&& -  && -&\\
		\hline\hline
	\end{tabular}
\end{table}

The low-lying spectra of $^{112}$Ba, $^{114}$Ba, $^{116}$Ce, $^{118}$Ce, and $^{120}$Ce are not known experimentally.
The SM $M$-scheme dimensions of them are around $2 \times 10^{10}$, $2 \times 10^{11}$, $2 \times 10^{12}$, $1 \times 10^{13}$, and $5 \times 10^{13}$, respectively, 
at the edge or beyond the reach of what the modern large-scale SM can do.
We calculate these five nuclei using the PNBCS and PHF. 
The calculated results are presented in Fig. \ref{fig:ba} and Table \ref{table5}, and the results of the $SDG$- and $SDGI$-pair approximations for $^{112,114}$Ba \cite{npadeformed3} are also included.
Coincidently, the excitation energies of $^{112,114}$Ba obtained by the PNBCS and $SDG$ are close to each other, and those obtained by the PHF and $SDGI$ are close to each other.
The $B(E2)$ values of $^{112,114}$Ba obtained by the PNBCS, PHF, and $SDGI$ are close to each others, while the $SDG$ result is $\sim 8\%$ smaller.

\section{summary}

In this work, we propose a simple approach to study the pair truncation of the shell model (SM) for rotational nuclei:
the projected number conserved BCS (PNBCS).
We generate deformed single-particle states by the HF calculation with Kramers degeneracy in a shell model basis, generalize the number conserved BCS to the case of open-shell nuclei, and project out collective states of good angular momentum using the linear algebraic projection approach. 
We study the shape evolution in the case of a 6-proton, 6-neutron system in the $pf$ shell by changing the relative strength of the
schematic pairing and quadrupole-quadrupole interactions.
We find the PNBCS provides a good description for low-lying states of well deformed and transitional nuclei. 

Applying the PNBCS to calculations of medium- and heavy-mass nuclei with SM effective interactions, 
we find that the low-lying level energies and $B(E2)$ values of the rotational bands are well reproduced by the PNBCS.
In particular our study of shape coexistence show that while the pair correlation and the configuration mixing between two different intrinsic states play a key role in reproducing the collective feature of the ground and side rotational bands in $^{66}$Ge and $^{68}$Ge, they are not important in $^{68}$Se.
Finally we point out our PNBCS prediction of low-lying states of $^{112,114}$Ba and $^{116,118,120}$Ce, which are beyond or almost beyond the capability of full-scale configuration-interaction SM.

Since the PNBCS computation is fast, it can be a powerful truncation scheme of the SM to study well deformed heavy-mass nuclei in a space larger than one major shell, e.g., the so-called extended extruded-intruded space.
Considering the variation after angular momentum projection and the number conserved Hartree-Fock-Bogoliubov are expected to further improve the results.
The generalization with broken-pair configurations can be a useful tool to study the phenomena of backbanding and nuclear shape-phase transition in heavy nuclei.

\begin{acknowledgments}
	This material is based upon work supported by
	the National Key R\&D Program of China under Grant No. 2018YFA0404403,
	the National Natural Science Foundation of China under Grant Nos. 12075169, 11705100, 12035011, 11975167, and 12175115, 
	the U.S. Department of Energy, Office of Science, Office of Nuclear Physics, under Award No. DE-FG02-03ER41272,	
	the CUSTIPEN (China-U.S. Theory Institute for Physics with Exotic Nuclei) funded by the U.S. Department of Energy, Office of Science grant number DE-SC0009971.
	Y. Lu is grateful for the support by the Youth Innovations and Talents Project of Shandong Provincial Colleges and Universities (Grant No. 201909118) and Higher Educational Youth Innovation Science and Technology Program Shandong Province (Grant No. 2020KJJ004).
\end{acknowledgments}

\vspace{0.2in}
\appendix{\bf APPENDIX: Formulas of overlaps between pair condensates and average energy of Hamiltonian}\label{appendix}

One refers to Ref. \cite{PhysRevC.99.014302} for the following formulas.
The normalization factors $\chi_{\tau,N}$ and $\chi^{[\alpha]}_{\tau,N}$ are calculated in the following recursive formulae:
\begin{eqnarray}
 &\displaystyle  \chi_{\tau,N}= N  \sum_{\alpha \in O_\tau} v_{\tau , \alpha}^2  \chi^{[\alpha]}_{\tau,N-1}, & \label{appendix_1}  \\
&\displaystyle	 \chi_{\tau,N} - \chi^{[\alpha]}_{\tau,N} = (N  v_{\tau , \alpha})^2 \chi^{[\alpha]}_{\tau,N-1} =  \chi_{\tau,N} \braket{\phi_{\tau,N} | \hat{n}_\alpha |\phi_{\tau,N}},  & \label{appendix_2}  \nonumber\\
\end{eqnarray}
With the initial values $\chi^{[\alpha]}_{\tau,N=0}=1$, one obtains $\chi_{\tau,N}$ by Eq. (\ref{appendix_1}) and then $\chi^{[\alpha]}_{\tau,N}$ by Eq. (\ref{appendix_2}).
$\chi^{[\alpha \beta]}_{\tau,N}$ and $\chi^{[\alpha \beta \gamma]}_{\tau,N}$ are successively calculated in the following recursive formulae:
\begin{eqnarray}
	& \chi^{[\beta]}_{\tau,N}-  \chi^{[\alpha \beta]}_{\tau,N} = (N  v_{\tau , \alpha})^2 \chi^{[\alpha \beta]}_{\tau,N-1}
	= \chi^{[\beta]}_{\tau,N} \braket{\phi^{[\beta]}_{\tau,N} | \hat{n}_\alpha |\phi^{[\beta]}_{\tau,N}},& \label{appendix_4} \nonumber\\ \\
&	\chi^{[\beta \gamma]}_{\tau,N}-  \chi^{[\alpha \beta \gamma]}_{\tau,N} = (N  v_{\tau , \alpha})^2 \chi^{[\alpha \beta \gamma]}_{\tau,N-1} 
	=\chi^{[\beta \gamma]}_{\tau,N} \braket{\phi^{[\beta \gamma ]}_{\tau,N} | \hat{n}_\alpha |\phi^{[\beta \gamma]}_{\tau,N}},& \label{appendix_6} \nonumber\\
\end{eqnarray}
with the initial values $\chi^{[\alpha \beta]}_{\tau,N=0}=\chi^{[\alpha \beta \gamma]}_{\tau,N=0}=1$.

Using Eq. (\ref{appendix_2}), the expectation value of an one-body operator, $ a^\dagger_{\tau,\alpha } a_{\tau,\beta}$, for the $N$-pair condensate is given by 
\begin{eqnarray}\label{appendix_7}
&& \braket{\phi_{\tau,N} |a^\dagger_{\tau,\alpha } a_{\tau,\beta}  |\phi_{\tau,N}} \nonumber\\
&  & \qquad   = \delta_{\alpha \beta} \braket{\phi_{\tau,N} | \hat{n}_\alpha  |\phi_{\tau,N}} 
= \delta_{\alpha \beta} \left(1- \frac{\chi^{[\alpha]}_{\tau,N}}{ \chi_{\tau,N} }\right) ,
\end{eqnarray}
which is nonvanishing if $\alpha  =\beta $.
For two-body operators $a^\dagger_{\tau,\alpha }  a^\dagger_{\tau,\beta } a_{\tau,\delta }  a_{\tau,\gamma }$, there are three types contribute nonvanishing values: 
$a^\dagger_{\tau,\alpha} a^\dagger_{\tau,\tilde{\alpha}} a_{\tau,\tilde{\alpha}} a_{\tau,\alpha} $, $a^\dagger_{\tau,\alpha} a^\dagger_{\tau,\tilde{\alpha}} a_{\tau,\tilde{\beta}} a_{\tau,\beta} $, and $a^\dagger_{\tau,\alpha} a^\dagger_{\tau,\beta}    a_{\tau,\beta}  a_{\tau,\alpha} $ ($\alpha \neq \beta$).
The expectation values are given by 

\begin{eqnarray}
 && \braket{\phi_{\tau,N} |  a^\dagger_{\tau,\alpha} a^\dagger_{\tau,\tilde{\alpha}} a_{\tau,\tilde{\alpha}} a_{\tau,\alpha}   |\phi_{\tau,N}} \nonumber\\
&  & \qquad \qquad \qquad   = \braket{\phi_{\tau,N} | \hat{n}_\alpha  |\phi_{\tau,N}} = 1- \frac{\chi^{[\alpha]}_{\tau,N}}{ \chi_{\tau,N} }  ,  \label{appendix_8} \\
&&	 \braket{\phi_{\tau,N} | a^\dagger_{\tau,\alpha} a^\dagger_{\tau,\tilde{\alpha}} a_{\tau,\tilde{\beta}} a_{\tau,\beta} |\phi_{\tau,N}} \nonumber\\
& & \qquad \qquad \qquad    = \braket{\phi_{\tau,N} | P^\dagger_{\tau,\alpha}  P_{\tau,\beta} |\phi_{\tau,N}} \nonumber\\
&&  \qquad \qquad \qquad    = N^2 v_{\tau,\alpha} v_{\tau,\beta}  \frac {	\chi^{[ \alpha \beta ]}_{\tau,N-1} } {\chi_{\tau,N} }  ,  \label{appendix_9} \\
&& \braket{\phi_{\tau,N} | a^\dagger_{\tau,\alpha} a^\dagger_{\tau,\beta}    a_{\tau,\beta}  a_{\tau,\alpha} |\phi_{\tau,N}} \nonumber\\
&  & \qquad \qquad \qquad    = \left(1 - \frac {\chi^{[\alpha]}_{\tau,N} + \chi^{[\beta]}_{\tau,N} - \chi^{[\alpha \beta]}_{\tau,N}}  {\chi_{\tau,N}}  \right),  \label{appendix_10}
\end{eqnarray}

\bibliography{bib,johnsonmaster,luyi-ref-papers}

\begin{thebibliography}{65}
\expandafter\ifx\csname natexlab\endcsname\relax\def\natexlab#1{#1}\fi
\expandafter\ifx\csname bibnamefont\endcsname\relax
  \def\bibnamefont#1{#1}\fi
\expandafter\ifx\csname bibfnamefont\endcsname\relax
  \def\bibfnamefont#1{#1}\fi
\expandafter\ifx\csname citenamefont\endcsname\relax
  \def\citenamefont#1{#1}\fi
\expandafter\ifx\csname url\endcsname\relax
  \def\url#1{\texttt{#1}}\fi
\expandafter\ifx\csname urlprefix\endcsname\relax\def\urlprefix{URL }\fi
\providecommand{\bibinfo}[2]{#2}
\providecommand{\eprint}[2][]{\url{#2}}

\bibitem[{\citenamefont{Talmi}(1993)}]{talmibook}
\bibinfo{author}{\bibfnamefont{I.}~\bibnamefont{Talmi}},
  \emph{\bibinfo{title}{Simple models of complex nuclei: the shell model and
  the interacting boson model}} (\bibinfo{publisher}{CRC Press},
  \bibinfo{year}{1993}).

\bibitem[{\citenamefont{Brown}(2001)}]{sm-brown}
\bibinfo{author}{\bibfnamefont{B.~A.} \bibnamefont{Brown}},
  \bibinfo{journal}{Progress in Particle and Nuclear Physics}
  \textbf{\bibinfo{volume}{47}}, \bibinfo{pages}{517} (\bibinfo{year}{2001}).

\bibitem[{\citenamefont{Talmi}(2005)}]{sm-talmi}
\bibinfo{author}{\bibfnamefont{I.}~\bibnamefont{Talmi}},
  \bibinfo{journal}{International Journal of Modern Physics E}
  \textbf{\bibinfo{volume}{14}}, \bibinfo{pages}{821} (\bibinfo{year}{2005}).

\bibitem[{\citenamefont{Caurier et~al.}(2005)\citenamefont{Caurier,
  Mart\'{\i}nez-Pinedo, Nowacki, Poves, and Zuker}}]{sm-caurier}
\bibinfo{author}{\bibfnamefont{E.}~\bibnamefont{Caurier}},
  \bibinfo{author}{\bibfnamefont{G.}~\bibnamefont{Mart\'{\i}nez-Pinedo}},
  \bibinfo{author}{\bibfnamefont{F.}~\bibnamefont{Nowacki}},
  \bibinfo{author}{\bibfnamefont{A.}~\bibnamefont{Poves}}, \bibnamefont{and}
  \bibinfo{author}{\bibfnamefont{A.~P.} \bibnamefont{Zuker}},
  \bibinfo{journal}{Rev. Mod. Phys.} \textbf{\bibinfo{volume}{77}},
  \bibinfo{pages}{427} (\bibinfo{year}{2005}).

\bibitem[{\citenamefont{Racah}(1942)}]{racah1942}
\bibinfo{author}{\bibfnamefont{G.}~\bibnamefont{Racah}},
  \bibinfo{journal}{Phys. Rev.} \textbf{\bibinfo{volume}{62}},
  \bibinfo{pages}{438} (\bibinfo{year}{1942}).

\bibitem[{\citenamefont{Racah}(1943)}]{racah1943}
\bibinfo{author}{\bibfnamefont{G.}~\bibnamefont{Racah}},
  \bibinfo{journal}{Phys. Rev.} \textbf{\bibinfo{volume}{63}},
  \bibinfo{pages}{367} (\bibinfo{year}{1943}).

\bibitem[{\citenamefont{Talmi}(1971)}]{talmi1}
\bibinfo{author}{\bibfnamefont{I.}~\bibnamefont{Talmi}},
  \bibinfo{journal}{Nucl. Phys. A} \textbf{\bibinfo{volume}{172}},
  \bibinfo{pages}{1} (\bibinfo{year}{1971}).

\bibitem[{\citenamefont{Shlomo and Talmi}(1972)}]{talmi2}
\bibinfo{author}{\bibfnamefont{S.}~\bibnamefont{Shlomo}} \bibnamefont{and}
  \bibinfo{author}{\bibfnamefont{I.}~\bibnamefont{Talmi}},
  \bibinfo{journal}{Nucl. Phys. A} \textbf{\bibinfo{volume}{198}},
  \bibinfo{pages}{81} (\bibinfo{year}{1972}).

\bibitem[{\citenamefont{Gambhir et~al.}(1969)\citenamefont{Gambhir, Rimini, and
  Weber}}]{brokenpair1}
\bibinfo{author}{\bibfnamefont{Y.~K.} \bibnamefont{Gambhir}},
  \bibinfo{author}{\bibfnamefont{A.}~\bibnamefont{Rimini}}, \bibnamefont{and}
  \bibinfo{author}{\bibfnamefont{T.}~\bibnamefont{Weber}},
  \bibinfo{journal}{Phys. Rev.} \textbf{\bibinfo{volume}{188}},
  \bibinfo{pages}{1573} (\bibinfo{year}{1969}).

\bibitem[{\citenamefont{Gambhir et~al.}(1981)\citenamefont{Gambhir, Haq, and
  Suri}}]{brokenpair2}
\bibinfo{author}{\bibfnamefont{Y.~K.} \bibnamefont{Gambhir}},
  \bibinfo{author}{\bibfnamefont{S.}~\bibnamefont{Haq}}, \bibnamefont{and}
  \bibinfo{author}{\bibfnamefont{J.~K.} \bibnamefont{Suri}},
  \bibinfo{journal}{Annals of Physics} \textbf{\bibinfo{volume}{133}},
  \bibinfo{pages}{154} (\bibinfo{year}{1981}).

\bibitem[{\citenamefont{Allaart et~al.}(1988)\citenamefont{Allaart, Boeker,
  Bonsignori, Savoai, and Gambhir}}]{brokenpair3}
\bibinfo{author}{\bibfnamefont{K.}~\bibnamefont{Allaart}},
  \bibinfo{author}{\bibfnamefont{E.}~\bibnamefont{Boeker}},
  \bibinfo{author}{\bibfnamefont{G.}~\bibnamefont{Bonsignori}},
  \bibinfo{author}{\bibfnamefont{M.}~\bibnamefont{Savoai}}, \bibnamefont{and}
  \bibinfo{author}{\bibfnamefont{Y.~K.} \bibnamefont{Gambhir}},
  \bibinfo{journal}{Physics Reports} \textbf{\bibinfo{volume}{169}},
  \bibinfo{pages}{209} (\bibinfo{year}{1988}).

\bibitem[{\citenamefont{Arima and Iachello}(1975)}]{ibm1}
\bibinfo{author}{\bibfnamefont{A.}~\bibnamefont{Arima}} \bibnamefont{and}
  \bibinfo{author}{\bibfnamefont{F.}~\bibnamefont{Iachello}},
  \bibinfo{journal}{Phys. Rev. Lett.} \textbf{\bibinfo{volume}{35}},
  \bibinfo{pages}{1069} (\bibinfo{year}{1975}).

\bibitem[{\citenamefont{Arima and Iachello}(1978)}]{ibm2}
\bibinfo{author}{\bibfnamefont{A.}~\bibnamefont{Arima}} \bibnamefont{and}
  \bibinfo{author}{\bibfnamefont{F.}~\bibnamefont{Iachello}},
  \bibinfo{journal}{Annals of Physics} \textbf{\bibinfo{volume}{111}},
  \bibinfo{pages}{201} (\bibinfo{year}{1978}).

\bibitem[{\citenamefont{Arima and Iachello}(1984)}]{ibm3}
\bibinfo{author}{\bibfnamefont{A.}~\bibnamefont{Arima}} \bibnamefont{and}
  \bibinfo{author}{\bibfnamefont{F.}~\bibnamefont{Iachello}},
  \emph{\bibinfo{title}{The Interacting Boson Model}}
  (\bibinfo{publisher}{Springer US}, \bibinfo{address}{Boston, MA},
  \bibinfo{year}{1984}).

\bibitem[{\citenamefont{Iachello and Arima}(1987)}]{ibm4}
\bibinfo{author}{\bibfnamefont{F.}~\bibnamefont{Iachello}} \bibnamefont{and}
  \bibinfo{author}{\bibfnamefont{A.}~\bibnamefont{Arima}},
  \emph{\bibinfo{title}{The Interacting Boson Model}}, Cambridge Monographs on
  Mathematical Physics (\bibinfo{publisher}{Cambridge University Press},
  \bibinfo{year}{1987}).

\bibitem[{\citenamefont{Chen}(1997)}]{npa3}
\bibinfo{author}{\bibfnamefont{J.-Q.} \bibnamefont{Chen}},
  \bibinfo{journal}{Nucl. Phys. A} \textbf{\bibinfo{volume}{626}},
  \bibinfo{pages}{686} (\bibinfo{year}{1997}).

\bibitem[{\citenamefont{Zhao et~al.}(2000)\citenamefont{Zhao, Yoshinaga,
  Yamaji, Chen, and Arima}}]{npaodd}
\bibinfo{author}{\bibfnamefont{Y.~M.} \bibnamefont{Zhao}},
  \bibinfo{author}{\bibfnamefont{N.}~\bibnamefont{Yoshinaga}},
  \bibinfo{author}{\bibfnamefont{S.}~\bibnamefont{Yamaji}},
  \bibinfo{author}{\bibfnamefont{J.~Q.} \bibnamefont{Chen}}, \bibnamefont{and}
  \bibinfo{author}{\bibfnamefont{A.}~\bibnamefont{Arima}},
  \bibinfo{journal}{Phys. Rev. C} \textbf{\bibinfo{volume}{62}},
  \bibinfo{pages}{014304} (\bibinfo{year}{2000}).

\bibitem[{\citenamefont{Zhao and Arima}(2014)}]{npareports}
\bibinfo{author}{\bibfnamefont{Y.~M.} \bibnamefont{Zhao}} \bibnamefont{and}
  \bibinfo{author}{\bibfnamefont{A.}~\bibnamefont{Arima}},
  \bibinfo{journal}{Physics Reports} \textbf{\bibinfo{volume}{545}},
  \bibinfo{pages}{1} (\bibinfo{year}{2014}).

\bibitem[{\citenamefont{Fu and Johnson}(2021)}]{npadeformed3}
\bibinfo{author}{\bibfnamefont{G.~J.} \bibnamefont{Fu}} \bibnamefont{and}
  \bibinfo{author}{\bibfnamefont{C.~W.} \bibnamefont{Johnson}},
  \bibinfo{journal}{Phys. Rev. C} \textbf{\bibinfo{volume}{104}},
  \bibinfo{pages}{024312} (\bibinfo{year}{2021}).

\bibitem[{\citenamefont{Fu and Johnson}(2020)}]{npadeformed1}
\bibinfo{author}{\bibfnamefont{G.~J.} \bibnamefont{Fu}} \bibnamefont{and}
  \bibinfo{author}{\bibfnamefont{C.~W.} \bibnamefont{Johnson}},
  \bibinfo{journal}{Physics Letters B} \textbf{\bibinfo{volume}{809}},
  \bibinfo{pages}{135705} (\bibinfo{year}{2020}).

\bibitem[{\citenamefont{Fu et~al.}(2021)\citenamefont{Fu, Johnson, Van~Isacker,
  and Ren}}]{npadeformed2}
\bibinfo{author}{\bibfnamefont{G.~J.} \bibnamefont{Fu}},
  \bibinfo{author}{\bibfnamefont{C.~W.} \bibnamefont{Johnson}},
  \bibinfo{author}{\bibfnamefont{P.}~\bibnamefont{Van~Isacker}},
  \bibnamefont{and} \bibinfo{author}{\bibfnamefont{Z.}~\bibnamefont{Ren}},
  \bibinfo{journal}{Phys. Rev. C} \textbf{\bibinfo{volume}{103}},
  \bibinfo{pages}{L021302} (\bibinfo{year}{2021}).

\bibitem[{\citenamefont{He et~al.}(2020)\citenamefont{He, Li, Luo, Zhang, Pan,
  and Draayer}}]{npam1}
\bibinfo{author}{\bibfnamefont{B.~C.} \bibnamefont{He}},
  \bibinfo{author}{\bibfnamefont{L.}~\bibnamefont{Li}},
  \bibinfo{author}{\bibfnamefont{Y.~A.} \bibnamefont{Luo}},
  \bibinfo{author}{\bibfnamefont{Y.}~\bibnamefont{Zhang}},
  \bibinfo{author}{\bibfnamefont{F.}~\bibnamefont{Pan}}, \bibnamefont{and}
  \bibinfo{author}{\bibfnamefont{J.~P.} \bibnamefont{Draayer}},
  \bibinfo{journal}{Phys. Rev. C} \textbf{\bibinfo{volume}{102}},
  \bibinfo{pages}{024304} (\bibinfo{year}{2020}).

\bibitem[{\citenamefont{Lei et~al.}(2021)\citenamefont{Lei, Lu, and
  Zhao}}]{npam2}
\bibinfo{author}{\bibfnamefont{Y.}~\bibnamefont{Lei}},
  \bibinfo{author}{\bibfnamefont{Y.}~\bibnamefont{Lu}}, \bibnamefont{and}
  \bibinfo{author}{\bibfnamefont{Y.~M.} \bibnamefont{Zhao}},
  \bibinfo{journal}{Chinese Physics C} \textbf{\bibinfo{volume}{45}},
  \bibinfo{pages}{054103} (\bibinfo{year}{2021}).

\bibitem[{\citenamefont{Bardeen
  et~al.}(1957{\natexlab{a}})\citenamefont{Bardeen, Cooper, and
  Schrieffer}}]{bcs1}
\bibinfo{author}{\bibfnamefont{J.}~\bibnamefont{Bardeen}},
  \bibinfo{author}{\bibfnamefont{L.~N.} \bibnamefont{Cooper}},
  \bibnamefont{and} \bibinfo{author}{\bibfnamefont{J.~R.}
  \bibnamefont{Schrieffer}}, \bibinfo{journal}{Phys. Rev.}
  \textbf{\bibinfo{volume}{106}}, \bibinfo{pages}{162}
  (\bibinfo{year}{1957}{\natexlab{a}}).

\bibitem[{\citenamefont{Bardeen
  et~al.}(1957{\natexlab{b}})\citenamefont{Bardeen, Cooper, and
  Schrieffer}}]{bcs2}
\bibinfo{author}{\bibfnamefont{J.}~\bibnamefont{Bardeen}},
  \bibinfo{author}{\bibfnamefont{L.~N.} \bibnamefont{Cooper}},
  \bibnamefont{and} \bibinfo{author}{\bibfnamefont{J.~R.}
  \bibnamefont{Schrieffer}}, \bibinfo{journal}{Phys. Rev.}
  \textbf{\bibinfo{volume}{108}}, \bibinfo{pages}{1175}
  (\bibinfo{year}{1957}{\natexlab{b}}).

\bibitem[{\citenamefont{Bohr et~al.}(1958)\citenamefont{Bohr, Mottelson, and
  Pines}}]{bcs-bohr}
\bibinfo{author}{\bibfnamefont{A.}~\bibnamefont{Bohr}},
  \bibinfo{author}{\bibfnamefont{B.~R.} \bibnamefont{Mottelson}},
  \bibnamefont{and} \bibinfo{author}{\bibfnamefont{D.}~\bibnamefont{Pines}},
  \bibinfo{journal}{Phys. Rev.} \textbf{\bibinfo{volume}{110}},
  \bibinfo{pages}{936} (\bibinfo{year}{1958}).

\bibitem[{\citenamefont{Migdal}(1959)}]{bcs-migdal}
\bibinfo{author}{\bibfnamefont{A.~B.} \bibnamefont{Migdal}},
  \bibinfo{journal}{Nuclear Physics} \textbf{\bibinfo{volume}{13}},
  \bibinfo{pages}{655} (\bibinfo{year}{1959}).

\bibitem[{\citenamefont{Belyaev}(1959)}]{bcs-belyaev}
\bibinfo{author}{\bibfnamefont{S.~T.} \bibnamefont{Belyaev}},
  \bibinfo{journal}{Kgl. Danske Videnskab. Selskab. Mat.-Fys. Medd.}
  \textbf{\bibinfo{volume}{31}} (\bibinfo{year}{1959}).

\bibitem[{\citenamefont{Ring and Schuck}(2004)}]{peterring}
\bibinfo{author}{\bibfnamefont{P.}~\bibnamefont{Ring}} \bibnamefont{and}
  \bibinfo{author}{\bibfnamefont{P.}~\bibnamefont{Schuck}},
  \emph{\bibinfo{title}{The nuclear many-body problem}}
  (\bibinfo{publisher}{Springer Science \& Business Media},
  \bibinfo{year}{2004}).

\bibitem[{\citenamefont{Bohr and Mottelson}(1998)}]{bohr1998nuclear2}
\bibinfo{author}{\bibfnamefont{A.}~\bibnamefont{Bohr}} \bibnamefont{and}
  \bibinfo{author}{\bibfnamefont{B.~R.} \bibnamefont{Mottelson}},
  \emph{\bibinfo{title}{Nuclear structure}}, vol.~\bibinfo{volume}{2}
  (\bibinfo{publisher}{World Scientific}, \bibinfo{year}{1998}).

\bibitem[{\citenamefont{Dietrich et~al.}(1964)\citenamefont{Dietrich, Mang, and
  Pradal}}]{Dietrich1964}
\bibinfo{author}{\bibfnamefont{K.}~\bibnamefont{Dietrich}},
  \bibinfo{author}{\bibfnamefont{H.~J.} \bibnamefont{Mang}}, \bibnamefont{and}
  \bibinfo{author}{\bibfnamefont{J.~H.} \bibnamefont{Pradal}},
  \bibinfo{journal}{Phys. Rev.} \textbf{\bibinfo{volume}{135}},
  \bibinfo{pages}{B22} (\bibinfo{year}{1964}).

\bibitem[{\citenamefont{Hara and Sun}(1995)}]{psm}
\bibinfo{author}{\bibfnamefont{K.}~\bibnamefont{Hara}} \bibnamefont{and}
  \bibinfo{author}{\bibfnamefont{Y.}~\bibnamefont{Sun}},
  \bibinfo{journal}{International Journal of Modern Physics E}
  \textbf{\bibinfo{volume}{04}}, \bibinfo{pages}{637} (\bibinfo{year}{1995}).

\bibitem[{\citenamefont{Chen et~al.}(2008)\citenamefont{Chen, Sun, and
  Gao}}]{PhysRevC.77.061305}
\bibinfo{author}{\bibfnamefont{Y.-S.} \bibnamefont{Chen}},
  \bibinfo{author}{\bibfnamefont{Y.}~\bibnamefont{Sun}}, \bibnamefont{and}
  \bibinfo{author}{\bibfnamefont{Z.-C.} \bibnamefont{Gao}},
  \bibinfo{journal}{Phys. Rev. C} \textbf{\bibinfo{volume}{77}},
  \bibinfo{pages}{061305} (\bibinfo{year}{2008}).

\bibitem[{\citenamefont{Dong et~al.}(2013)\citenamefont{Dong, Wang, Liu, and
  Xu}}]{PhysRevC.88.024328}
\bibinfo{author}{\bibfnamefont{G.~X.} \bibnamefont{Dong}},
  \bibinfo{author}{\bibfnamefont{X.~B.} \bibnamefont{Wang}},
  \bibinfo{author}{\bibfnamefont{H.~L.} \bibnamefont{Liu}}, \bibnamefont{and}
  \bibinfo{author}{\bibfnamefont{F.~R.} \bibnamefont{Xu}},
  \bibinfo{journal}{Phys. Rev. C} \textbf{\bibinfo{volume}{88}},
  \bibinfo{pages}{024328} (\bibinfo{year}{2013}).

\bibitem[{\citenamefont{Jia}(2017)}]{PhysRevC.96.034313}
\bibinfo{author}{\bibfnamefont{L.~Y.} \bibnamefont{Jia}},
  \bibinfo{journal}{Phys. Rev. C} \textbf{\bibinfo{volume}{96}},
  \bibinfo{pages}{034313} (\bibinfo{year}{2017}).

\bibitem[{\citenamefont{Jia}(2019)}]{PhysRevC.99.014302}
\bibinfo{author}{\bibfnamefont{L.~Y.} \bibnamefont{Jia}},
  \bibinfo{journal}{Phys. Rev. C} \textbf{\bibinfo{volume}{99}},
  \bibinfo{pages}{014302} (\bibinfo{year}{2019}).

\bibitem[{\citenamefont{Schmid}(2004)}]{SCHMID2004565}
\bibinfo{author}{\bibfnamefont{K.~W.} \bibnamefont{Schmid}},
  \bibinfo{journal}{Progress in Particle and Nuclear Physics}
  \textbf{\bibinfo{volume}{52}}, \bibinfo{pages}{565} (\bibinfo{year}{2004}).

\bibitem[{\citenamefont{Gao et~al.}(2015)\citenamefont{Gao, Horoi, and
  Chen}}]{PhysRevC.92.064310}
\bibinfo{author}{\bibfnamefont{Z.-C.} \bibnamefont{Gao}},
  \bibinfo{author}{\bibfnamefont{M.}~\bibnamefont{Horoi}}, \bibnamefont{and}
  \bibinfo{author}{\bibfnamefont{Y.~S.} \bibnamefont{Chen}},
  \bibinfo{journal}{Phys. Rev. C} \textbf{\bibinfo{volume}{92}},
  \bibinfo{pages}{064310} (\bibinfo{year}{2015}).

\bibitem[{\citenamefont{Lu et~al.}(2022{\natexlab{a}})\citenamefont{Lu, Lei,
  Johnson, and Shen}}]{PhysRevC.105.034317}
\bibinfo{author}{\bibfnamefont{Y.}~\bibnamefont{Lu}},
  \bibinfo{author}{\bibfnamefont{Y.}~\bibnamefont{Lei}},
  \bibinfo{author}{\bibfnamefont{C.~W.} \bibnamefont{Johnson}},
  \bibnamefont{and} \bibinfo{author}{\bibfnamefont{J.~J.} \bibnamefont{Shen}},
  \bibinfo{journal}{Phys. Rev. C} \textbf{\bibinfo{volume}{105}},
  \bibinfo{pages}{034317} (\bibinfo{year}{2022}{\natexlab{a}}).

\bibitem[{\citenamefont{Lu et~al.}(2022{\natexlab{b}})\citenamefont{Lu, Lei,
  Johnson, and Shen}}]{lu2022nuclear}
\bibinfo{author}{\bibfnamefont{Y.}~\bibnamefont{Lu}},
  \bibinfo{author}{\bibfnamefont{Y.}~\bibnamefont{Lei}},
  \bibinfo{author}{\bibfnamefont{C.~W.} \bibnamefont{Johnson}},
  \bibnamefont{and} \bibinfo{author}{\bibfnamefont{J.~J.} \bibnamefont{Shen}}
  (\bibinfo{year}{2022}{\natexlab{b}}), \eprint{arXiv.2112.15393}.

\bibitem[{\citenamefont{Johnson and O'Mara}(2017)}]{PHF1}
\bibinfo{author}{\bibfnamefont{C.~W.} \bibnamefont{Johnson}} \bibnamefont{and}
  \bibinfo{author}{\bibfnamefont{K.~D.} \bibnamefont{O'Mara}},
  \bibinfo{journal}{Phys. Rev. C} \textbf{\bibinfo{volume}{96}},
  \bibinfo{pages}{064304} (\bibinfo{year}{2017}).

\bibitem[{\citenamefont{Kramers}(1930)}]{kramers}
\bibinfo{author}{\bibfnamefont{H.~A.} \bibnamefont{Kramers}},
  \bibinfo{journal}{Proceedings of the Royal Netherlands Academy of Arts and
  Sciences} \textbf{\bibinfo{volume}{33}}, \bibinfo{pages}{959}
  (\bibinfo{year}{1930}), ISSN \bibinfo{issn}{0370-0348}.

\bibitem[{\citenamefont{Jia}(2015)}]{jia2015practical}
\bibinfo{author}{\bibfnamefont{L.~Y.} \bibnamefont{Jia}},
  \bibinfo{journal}{Journal of Physics G: Nuclear and Particle Physics}
  \textbf{\bibinfo{volume}{42}}, \bibinfo{pages}{115105}
  (\bibinfo{year}{2015}).

\bibitem[{\citenamefont{Edmonds}(1957)}]{Edmonds1957}
\bibinfo{author}{\bibfnamefont{A.~R.} \bibnamefont{Edmonds}},
  \emph{\bibinfo{title}{Angular momentum in quantum mechanics}}
  (\bibinfo{publisher}{Princeton University Press}, \bibinfo{year}{1957}).

\bibitem[{\citenamefont{Johnson and Jiao}(2019)}]{PHF2}
\bibinfo{author}{\bibfnamefont{C.~W.} \bibnamefont{Johnson}} \bibnamefont{and}
  \bibinfo{author}{\bibfnamefont{C.}~\bibnamefont{Jiao}}, \bibinfo{journal}{J.
  Phys. G: Nucl. Part. Phys.} \textbf{\bibinfo{volume}{46}},
  \bibinfo{pages}{015101} (\bibinfo{year}{2019}).

\bibitem[{\citenamefont{Poves et~al.}(2001)\citenamefont{Poves,
  Sánchez-Solano, Caurier, and Nowacki}}]{POVES2001157}
\bibinfo{author}{\bibfnamefont{A.}~\bibnamefont{Poves}},
  \bibinfo{author}{\bibfnamefont{J.}~\bibnamefont{Sánchez-Solano}},
  \bibinfo{author}{\bibfnamefont{E.}~\bibnamefont{Caurier}}, \bibnamefont{and}
  \bibinfo{author}{\bibfnamefont{F.}~\bibnamefont{Nowacki}},
  \bibinfo{journal}{Nuclear Physics A} \textbf{\bibinfo{volume}{694}},
  \bibinfo{pages}{157} (\bibinfo{year}{2001}).

\bibitem[{\citenamefont{Johnson et~al.}(2013)\citenamefont{Johnson, Ormand, and
  Krastev}}]{bigstick}
\bibinfo{author}{\bibfnamefont{C.~W.} \bibnamefont{Johnson}},
  \bibinfo{author}{\bibfnamefont{W.~E.} \bibnamefont{Ormand}},
  \bibnamefont{and} \bibinfo{author}{\bibfnamefont{P.~G.}
  \bibnamefont{Krastev}}, \bibinfo{journal}{Computer Physics Communications}
  \textbf{\bibinfo{volume}{184}}, \bibinfo{pages}{2761} (\bibinfo{year}{2013}).

\bibitem[{\citenamefont{Johnson et~al.}(2018)\citenamefont{Johnson, Ormand,
  McElvain, and Shan}}]{johnson2018bigstick}
\bibinfo{author}{\bibfnamefont{C.~W.} \bibnamefont{Johnson}},
  \bibinfo{author}{\bibfnamefont{W.~E.} \bibnamefont{Ormand}},
  \bibinfo{author}{\bibfnamefont{K.~S.} \bibnamefont{McElvain}},
  \bibnamefont{and} \bibinfo{author}{\bibfnamefont{H.}~\bibnamefont{Shan}},
  \bibinfo{journal}{arXiv preprint arXiv:1801.08432}  (\bibinfo{year}{2018}).

\bibitem[{\citenamefont{Wu}(2000)}]{WU20001}
\bibinfo{author}{\bibfnamefont{S.-C.} \bibnamefont{Wu}},
  \bibinfo{journal}{Nuclear Data Sheets} \textbf{\bibinfo{volume}{91}},
  \bibinfo{pages}{1} (\bibinfo{year}{2000}).

\bibitem[{\citenamefont{Burrows}(2006)}]{BURROWS20061747}
\bibinfo{author}{\bibfnamefont{T.~W.} \bibnamefont{Burrows}},
  \bibinfo{journal}{Nuclear Data Sheets} \textbf{\bibinfo{volume}{107}},
  \bibinfo{pages}{1747} (\bibinfo{year}{2006}).

\bibitem[{\citenamefont{Chen et~al.}(2011)\citenamefont{Chen, Singh, and
  Cameron}}]{CHEN20112357}
\bibinfo{author}{\bibfnamefont{J.}~\bibnamefont{Chen}},
  \bibinfo{author}{\bibfnamefont{B.}~\bibnamefont{Singh}}, \bibnamefont{and}
  \bibinfo{author}{\bibfnamefont{J.~A.} \bibnamefont{Cameron}},
  \bibinfo{journal}{Nuclear Data Sheets} \textbf{\bibinfo{volume}{112}},
  \bibinfo{pages}{2357} (\bibinfo{year}{2011}).

\bibitem[{\citenamefont{Chen and Singh}(2019)}]{CHEN20191}
\bibinfo{author}{\bibfnamefont{J.}~\bibnamefont{Chen}} \bibnamefont{and}
  \bibinfo{author}{\bibfnamefont{B.}~\bibnamefont{Singh}},
  \bibinfo{journal}{Nuclear Data Sheets} \textbf{\bibinfo{volume}{157}},
  \bibinfo{pages}{1} (\bibinfo{year}{2019}).

\bibitem[{\citenamefont{Dong and Junde}(2015)}]{DONG2015185}
\bibinfo{author}{\bibfnamefont{Y.}~\bibnamefont{Dong}} \bibnamefont{and}
  \bibinfo{author}{\bibfnamefont{H.}~\bibnamefont{Junde}},
  \bibinfo{journal}{Nuclear Data Sheets} \textbf{\bibinfo{volume}{128}},
  \bibinfo{pages}{185} (\bibinfo{year}{2015}).

\bibitem[{\citenamefont{Browne and Tuli}(2013)}]{BROWNE20131849}
\bibinfo{author}{\bibfnamefont{E.}~\bibnamefont{Browne}} \bibnamefont{and}
  \bibinfo{author}{\bibfnamefont{J.~K.} \bibnamefont{Tuli}},
  \bibinfo{journal}{Nuclear Data Sheets} \textbf{\bibinfo{volume}{114}},
  \bibinfo{pages}{1849} (\bibinfo{year}{2013}).

\bibitem[{\citenamefont{Nichols et~al.}(2012)\citenamefont{Nichols, Singh, and
  Tuli}}]{NICHOLS2012973}
\bibinfo{author}{\bibfnamefont{A.~L.} \bibnamefont{Nichols}},
  \bibinfo{author}{\bibfnamefont{B.}~\bibnamefont{Singh}}, \bibnamefont{and}
  \bibinfo{author}{\bibfnamefont{J.~K.} \bibnamefont{Tuli}},
  \bibinfo{journal}{Nuclear Data Sheets} \textbf{\bibinfo{volume}{113}},
  \bibinfo{pages}{973} (\bibinfo{year}{2012}).

\bibitem[{\citenamefont{Singh}(2007)}]{SINGH2007197}
\bibinfo{author}{\bibfnamefont{B.}~\bibnamefont{Singh}},
  \bibinfo{journal}{Nuclear Data Sheets} \textbf{\bibinfo{volume}{108}},
  \bibinfo{pages}{197} (\bibinfo{year}{2007}).

\bibitem[{\citenamefont{Pritychenko et~al.}(2012)\citenamefont{Pritychenko,
  Choquette, Horoi, Karamy, and Singh}}]{PRITYCHENKO2012798}
\bibinfo{author}{\bibfnamefont{B.}~\bibnamefont{Pritychenko}},
  \bibinfo{author}{\bibfnamefont{J.}~\bibnamefont{Choquette}},
  \bibinfo{author}{\bibfnamefont{M.}~\bibnamefont{Horoi}},
  \bibinfo{author}{\bibfnamefont{B.}~\bibnamefont{Karamy}}, \bibnamefont{and}
  \bibinfo{author}{\bibfnamefont{B.}~\bibnamefont{Singh}},
  \bibinfo{journal}{Atomic Data and Nuclear Data Tables}
  \textbf{\bibinfo{volume}{98}}, \bibinfo{pages}{798} (\bibinfo{year}{2012}).

\bibitem[{\citenamefont{Browne and Tuli}(2010)}]{BROWNE20101093}
\bibinfo{author}{\bibfnamefont{E.}~\bibnamefont{Browne}} \bibnamefont{and}
  \bibinfo{author}{\bibfnamefont{J.~K.} \bibnamefont{Tuli}},
  \bibinfo{journal}{Nuclear Data Sheets} \textbf{\bibinfo{volume}{111}},
  \bibinfo{pages}{1093} (\bibinfo{year}{2010}).

\bibitem[{\citenamefont{McCutchan}(2012)}]{MCCUTCHAN20121735}
\bibinfo{author}{\bibfnamefont{E.~A.} \bibnamefont{McCutchan}},
  \bibinfo{journal}{Nuclear Data Sheets} \textbf{\bibinfo{volume}{113}},
  \bibinfo{pages}{1735} (\bibinfo{year}{2012}).

\bibitem[{\citenamefont{Honma et~al.}(2009)\citenamefont{Honma, Otsuka,
  Mizusaki, and Hjorth-Jensen}}]{PhysRevC.80.064323}
\bibinfo{author}{\bibfnamefont{M.}~\bibnamefont{Honma}},
  \bibinfo{author}{\bibfnamefont{T.}~\bibnamefont{Otsuka}},
  \bibinfo{author}{\bibfnamefont{T.}~\bibnamefont{Mizusaki}}, \bibnamefont{and}
  \bibinfo{author}{\bibfnamefont{M.}~\bibnamefont{Hjorth-Jensen}},
  \bibinfo{journal}{Phys. Rev. C} \textbf{\bibinfo{volume}{80}},
  \bibinfo{pages}{064323} (\bibinfo{year}{2009}).

\bibitem[{\citenamefont{Hjorth-Jensen et~al.}(1995)\citenamefont{Hjorth-Jensen,
  Kuo, and Osnes}}]{HJORTHJENSEN1995125}
\bibinfo{author}{\bibfnamefont{M.}~\bibnamefont{Hjorth-Jensen}},
  \bibinfo{author}{\bibfnamefont{T.~T.} \bibnamefont{Kuo}}, \bibnamefont{and}
  \bibinfo{author}{\bibfnamefont{E.}~\bibnamefont{Osnes}},
  \bibinfo{journal}{Physics Reports} \textbf{\bibinfo{volume}{261}},
  \bibinfo{pages}{125} (\bibinfo{year}{1995}).

\bibitem[{\citenamefont{Heyde and Wood}(2011)}]{Heyde-rmp}
\bibinfo{author}{\bibfnamefont{K.}~\bibnamefont{Heyde}} \bibnamefont{and}
  \bibinfo{author}{\bibfnamefont{J.~L.} \bibnamefont{Wood}},
  \bibinfo{journal}{Rev. Mod. Phys.} \textbf{\bibinfo{volume}{83}},
  \bibinfo{pages}{1467} (\bibinfo{year}{2011}).

\bibitem[{\citenamefont{Gürdal and Kondev}(2012)}]{GURDAL20121315}
\bibinfo{author}{\bibfnamefont{G.}~\bibnamefont{Gürdal}} \bibnamefont{and}
  \bibinfo{author}{\bibfnamefont{F.~G.} \bibnamefont{Kondev}},
  \bibinfo{journal}{Nuclear Data Sheets} \textbf{\bibinfo{volume}{113}},
  \bibinfo{pages}{1315} (\bibinfo{year}{2012}).

\bibitem[{\citenamefont{Qi}()}]{private1}
\bibinfo{author}{\bibfnamefont{C.}~\bibnamefont{Qi}},
  \bibinfo{howpublished}{private communications}.

\bibitem[{\citenamefont{Blachot}(2000)}]{BLACHOT2000135}
\bibinfo{author}{\bibfnamefont{J.}~\bibnamefont{Blachot}},
  \bibinfo{journal}{Nuclear Data Sheets} \textbf{\bibinfo{volume}{91}},
  \bibinfo{pages}{135} (\bibinfo{year}{2000}).

\end{thebibliography}

\end{document}